\documentclass[11pt,xcolor=dvipsnames]{article}
\DeclareMathAlphabet{\scr}{U}{rsfs}{m}{n}

\usepackage{scalerel,stackengine}
\usepackage[colorlinks,urlcolor=black,linkcolor = black,citecolor = black,]{hyperref}
\usepackage{subfigure}
\usepackage{latexsym}
\usepackage{epsfig}
\usepackage[mathscr]{eucal}
\usepackage{amsfonts}
\usepackage{amscd}
\usepackage{cite}
\usepackage{array}
\usepackage{amssymb}
\usepackage{colordvi}
\usepackage[centertags]{amsmath}
\usepackage{enumerate}
\usepackage{graphicx}
\usepackage{booktabs}
\usepackage{theorem}
\usepackage[footnotesize]{caption}
\usepackage{soul}
\usepackage{mcite}
\usepackage{slashed}
\usepackage{xcolor}
\usepackage{ulem}
\usepackage{bbm}
\usepackage[utf8]{inputenc}
\usepackage{fancyvrb}
\usepackage{framed}
\usepackage{xspace}
\newcommand{\newc}{\newcommand}
\newc{\be}{\begin{equation}}
\newc{\bea}{\begin{eqnarray}}
\newc{\eea}{\end{eqnarray}}
\newc{\ol}{\overline}
\newc{\wt}{\widetilde}
\newc{\bs}{\boldsymbol}
\newc{\m}{\mathcal}
\newc{\la}{\langle}
\newc{\ra}{\rangle}

\newcommand\equalhat{\mathrel{\stackon[1.5pt]{=}{\stretchto{%
    \scalerel*[\widthof{=}]{\wedge}{\rule{1ex}{3ex}}}{0.5ex}}}}

\newcommand{\beq}{\begin{eqnarray}}
\newcommand{\eeq}{\end{eqnarray}}
\newcommand{\bpmatrix}{\begin{pmatrix}}
\newcommand{\epmatrix}{\end{pmatrix}}
\newcommand{\ba}{\begin{array}}
\newcommand{\ea}{\end{array}}

\renewcommand{\ol}{\text{1l}}

\renewcommand{\Re}{\text{Re}\!}
\renewcommand{\Im}{\text{Im}\!}

\renewcommand{\eqref}[1]{Eq.~(\ref{#1})}

\newcommand{\bc}{\begin{center}}
\newcommand{\ec}{\end{center}}

\newcommand{\s}{\newline \vspace*{-3.5mm}}

\newcommand{\ii}{\mathrm{i}}

\newcommand{\summe}[2]{\sum\limits_{#1}^{#2}}

\newcommand{\bint}[2]{\int\limits_{#1}^{#2}}

\newcommand{\Tr}{\mathrm{Tr}}

\newcommand{\CW}{\text{CW}}

\newcommand{\dash}{{-}}

\newcommand{\figref}[2][{}]{\hyperref[#2]{\figurename~\ref{#2}#1}} 

\newenvironment{kasten*}[1]
{
\hspace{0.05\linewidth}
\begin{minipage}{0.95\linewidth}
\setlength{\fboxsep}{10pt}
\definecolor{shadecolor}{gray}{0.9}
\definecolor{framecolor}{gray}{0}

\MakeFramed {\FrameRestore}
\subsection*{#1}
}
{
\endMakeFramed
\end{minipage}
\vspace{1em}
}

\newcommand{\download}{\url{https://github.com/phbasler/BSMPT}}

\newcommand{\PathToYourLib}{PathToYourLib}
\newcommand{\home}{\$BSMPT}

\allowdisplaybreaks

\begin{document}
\title{
\vspace*{-3cm}
\phantom{h} \hfill\mbox{\small KA-TP-06-2018}
\\[1cm]
\textbf{BSMPT\\
Beyond the Standard Model Phase Transitions \\ A Tool for the
Electroweak Phase Transition in Extended Higgs Sectors\\[4mm]}} 

\date{}
\author{
Philipp Basler$^{1\,}$\footnote{E-mail:
  \texttt{philipp.basler@kit.edu}} ,
Margarete M\"{u}hlleitner$^{1\,}$\footnote{E-mail:
\texttt{margarete.muehlleitner@kit.edu}} 
\\[9mm]
{\small\it
$^1$Institute for Theoretical Physics, Karlsruhe Institute of Technology,} \\
{\small\it 76128 Karlsruhe, Germany}}

\maketitle

\begin{abstract}
We provide the {\tt C++} tool {\tt BSMPT} for calculating the strength
of the electroweak phase transition in extended Higgs sectors.  This
relies on the loop-corrected effective potential at finite temperature
including daisy resummation of the bosonic masses. The program
allows to compute the vacuum expectation value (VEV) $v$ of the potential
as a function of the temperature, and in particular the critical VEV
$v_c$ at the temperature $T_c$ where the phase transition takes
place. In addition, the loop-corrected trilinear Higgs self-couplings are
provided. We apply an 'on-shell' renormalization scheme in the sense
that the loop-corrected masses and mixing angles are required to be
equal to their tree-level input values. This allows for efficient
scans in the parameter space of the models. The models implemented so far
are the CP-conserving and the CP-violating 2-Higgs-Doublet Models (2HDM) and the
Next-to-Minimal 2HDM (N2HDM). The program structure is such that the
user can easily implement further models. Our tool can be used for the
investigation of electroweak baryogenesis in models with extended
Higgs sectors and the related Higgs self-couplings. The combination
with parameter scans in the respective models allows to study the
impact on collider phenomenology and to make a link between collider
phenomenology and cosmology. The program package can be downloaded at:
\download. 
\end{abstract}
\thispagestyle{empty}
\vfill
\newpage
\setcounter{page}{1}


\section{Introduction} 
The observed baryon asymmetry of the Universe (BAU)
\cite{Bennett:2012zja} is one of the unsolved puzzles within the
Standard Model (SM). Electroweak (EW) baryogenesis provides a
mechanism to generate the BAU dynamically in the early
Universe during a first order EW phase transition (EWPT)
\cite{Kuzmin:1985mm,Cohen:1990it,Cohen:1993nk,Quiros:1994dr,Rubakov:1996vz,Funakubo:1996dw,Trodden:1998ym,Bernreuther:2002uj,Morrissey:2012db}
provided all three Sakharov conditions \cite{sakharov} are
fulfilled. Although in the SM all three conditions can in 
principle be fulfilled, the phase transition (PT) is not of strong first order
\cite{Morrissey:2012db,smnot,notsm2}, so that new 
physics extensions are required that provide additional sources of
CP violation as well as further scalar states triggering a first order
EWPT. The investigation of the PT requires the computation of the
loop-corrected Higgs potential at finite temperature, in order find
the vacuum expectation value (VEV) $v_c$ at the critical temperature
$T_c$. The latter is defined as the temperature where two degenerate
global minima exist. A value of $\xi_c = v_c / T_c >1$ indicates a strong
first order PT \cite{Quiros:1994dr,Moore:1998swa}. \s

In this paper we present the program package
{\tt BSMPT} - 'Beyond the Standard Model Phase Transitions':
\begin{itemize}
\item[] A {\tt C++} tool for the calculation of the loop-corrected effective potential at
  finite temperature \cite{ColemanWeinberg,Quiros:1999jp,Dolan:1973qd}
  including the daisy resummation for the bosonic masses
  \cite{Carrington:1991hz}. The latter is included in two different
  approximations for the treatment of the thermal masses, the Parwani
  \cite{Parwani:1991gq} and the Arnold-Espinosa method
  \cite{Arnold:1992rz}, where the Arnold-Espinosa method is set as the
  default one. The renormalization of the potential is based on
  physical conditions. These are 'on-shell' conditions in the sense that the 
  loop-corrected masses and mixing angles extracted from the effective
  potential are forced to be equal to their tree-level input
  values. 
\end{itemize} 
The package can be used for:
\begin{itemize}
\item[-] The calculation of the EWPT: For a given point in the
  parameter space, it calculates the global minimum of the potential
  at a given temperature and determines the critical
  temperature $T_c$ where the phase transition takes place together with the
  corresponding VEV, $v_c$.\footnote{Note, that we do not consider the possibility of a
  2-state PT \cite{} in our models.} These two values are then used to compute 
  the strength of the PT, parametrized by $\xi_c= v_c/T_c$. 
\item[-] The calculation of the evolution of the
VEV(s)\footnote{In extended Higgs sectors we have several VEVs, which,
  at zero temperature, combine to the total VEV $v \approx 246.22$~GeV.}
with the temperature.
\item[-] The calculation of the global minimum of the 1-loop corrected
  potential at zero temperature. 
\item[-] The calculation of the loop-corrected trilinear Higgs
  self-couplings in the on-shell scheme. 
\end{itemize}
For the combined investigation of the PT through EW baryogenesis
together with collider phenomenology it is recommended to use input
parameter points that already fulfill all relevant experimental and
theoretical constraints in order to pin down the viable parameter
space as much as possible. Our chosen on-shell renormalization has the
advantage to allow for efficient scans in the parameter space of the
investigated models and simultaneously take into account all relevant
theoretical and up-to-date experimental constraints. For sample
applications, see Refs.~\cite{Basler:2016obg} and \cite{Basler:2017uxn} in
the CP-conserving and CP-violating 2-Higgs Doublet Model
(2HDM), respectively. \s  

The program was developed and tested on an OpenSuse 42.2, Ubuntu
14.04, Ubuntu 16.04 and Mac 10.13 system with {\tt g++
  v6.2.1} and {\tt g++ v.7.2.1}. The package can be downloaded at: 

\begin{center}
\download
\end{center}

The outline of the paper is as follows. In Section~\ref{sec:calc} we
present our calculation which also serves to set our notation. The
models that are already implemented in the package are introduced in
Section~\ref{sec:implmodels}. 
In Section~\ref{sec:install} we explain how to install
and run the program. 
Section~\ref{sec:executables} describes the available executables and
their corresponding output files. 
Section~\ref{sec:newmodels} explains
with the help of a toy model how a new model can be added to the
program package. The summary is given in Section~\ref{sec:concl}. \s


\section{Calculation \label{sec:calc}}
In order to investigate the properties of the EWPT, the loop-corrected
effective potential $V$ at finite temperature $T$ has to be
computed. In terms of the static field configuration $\omega$ and the
temperature $T$ the potential 
\beq
V = V(\omega,T)
\eeq 
develops a minimum for the ground state $\omega = v(T)$. 
In case $v= 0$ we are in the symmetric phase of the
model, for $v \ne 0$, we are in the broken
phase. Starting with the symmetric vacuum in the early universe, the
EWPT is defined as the point in the evolution of the
potential, where a second minimum with non-zero VEV $v_c$ 
developed at the critical temperature $T_c$, for which
\beq
V (v=0, T_c) = V (v=v_c, T_c) \;.
\eeq
The thermal evolution of the ground state of the potential is an
important criterion to judge the fulfillment of the Sakharov
criteria. In order to be a possible candidate for electroweak 
baryogenesis the EWPT has to be of strong first order, defined as
\cite{Quiros:1994dr,Moore:1998swa} 
\beq
\xi_c \equiv \frac{v_c}{T_c} > 1  \;.
\eeq
Because of the rich structure of the electroweak potential the
calculation of $v_c$ and $T_c$ is not possible in an analytic way and
we therefore present this program which calculates $v_c$ and $T_c$
numerically.  \s

The loop-corrected effective potential at finite temperature $T$ as
function of the classical constant field configuration, generically denoted by
$\omega$, reads
\beq
V (\omega,T) = V (\omega) + V^T (\omega,T) \equiv V^{(0)} (\omega) +
V^{\text{CW}} (\omega) + V^{\text{CT}} (\omega) + V^T (\omega,T) \;.
\label{eq:fielconfig}
\eeq
In $V(\omega)$ we summarize the contributions that do not depend
explicitly on the temperature $T$. These are the tree-level potential 
$V^{(0)}$, the Coleman-Weinberg potential $V^{\text{CW}}$ and the
coun\-ter\-term potential $V^{\text{CT}}$. The thermal corrections at finite
temperature $T$ are given by $V^T (\omega,T)$.

\subsection{Notation \label{sec:notation}}
We use the notation of Ref.~\cite{Camargo-Molina:2016moz}\footnote{The
  additional terms appearing in \cite{Camargo-Molina:2016moz} do not
  exist in our models and are therefore omitted here.} in which 
the tree-level Lagrangian, relevant for the effective potential, can be cast into the form 
\begin{align}
-\mathcal{L}_S &= L^i \Phi_i + \frac{1}{2!} L^{ij} \Phi_i \Phi_j+\frac{1}{3!} L^{ijk} \Phi_i
\Phi_j\Phi_k + \frac{1}{4!} L^{ijkl} \Phi_i \Phi_j \Phi_k \Phi_l \label{Eq:LS_Classical}\\
-\mathcal{L}_F &= \frac{1}{2} Y^{IJk} \Psi_I \Psi_J \Phi_k + c.c. \label{Eq:LF_Classical}\\
\mathcal{L}_{G} &= \frac{1}{4} G^{abij}
A_{a\mu}A_b^\mu\Phi_i\Phi_j \;,  \label{Eq:LG_Classical} 
\end{align}
for every model applied in the code. Here and in the
following we adopt the Einstein convention and sum over repeated
indices if one is up and the other down, otherwise not. In this description the 
scalar multiplets are decomposed into $n_{\text{Higgs}}$ real scalar
fields $\Phi_i$, with $i= 1,\dots , n_{\text{Higgs}}$. The
fermion multiplets are represented through $n_{\text{fermion}}$ Weyl
spinors $\Psi_I$, with $I=1,\dots , n_{\text{fermion}}$.  The gauge bosons are given by
the four-vectors $A_\mu^a$. The gauge group index $a$ runs over
$n_{\text{gauge}}$ gauge bosons in the adjoint representation of the
gauge group. The extended Higgs potential is
given by $-\mathcal{L}_S$ and is described through the tensors
$L^i,L^{ij},L^{ijk},L^{ijkl}$ and the real scalar fields $\Phi_i$
($i,j,k,l= 1, \dots , n_{\text{Higgs}}$). The
interactions between the scalar fields and the fermions $\Psi_I$ are
described by the tensor $Y^{IJk}$ ($I,J = 1\dots n_{\text{fermion}}$).  The
interactions between the scalars and the gauge
bosons $A_\mu^a$ are given by $G^{abij}$ ($a,b = 1\dots
n_{\text{gauge}}$). After symmetry breaking the scalar fields $\Phi_i$
are expanded around a classical constant field configuration $\omega_i$ as
\begin{align}
\Phi_i(x) &=  \omega_i + \phi_i(x) \;, \label{eq:dev}
\end{align}
where the $\phi_i(x)$ describe the quantum scalar field
fluctuations. After inserting Eq.~(\ref{eq:dev}) in
Eqs.~(\ref{Eq:LS_Classical})-(\ref{Eq:LG_Classical}), they can be
rewritten as 
\begin{align}
-\mathcal{L}_S &= \Lambda + \Lambda^i_{(S)} \phi_i + \frac{1}{2} \Lambda_{(S)}^{ij} \phi_i\phi_j + \frac{1}{3!} \Lambda^{ijk}_{(S)} \phi_i\phi_j\phi_k + \frac{1}{4!} \Lambda_{(S)}^{ijkl}\phi_i\phi_j\phi_k\phi_l \\
-\mathcal{L}_F &= \frac{1}{2} M^{IJ} \Psi_I\Psi_J + \frac{1}{2}
Y^{IJk}\Psi_I\Psi_J\phi_k + c.c.  \\
\mathcal{L}_G &= \frac{1}{2} \Lambda^{ab}_{(G)} A_{a\mu}A_b^{\mu}
+\frac{1}{2} \Lambda^{abi}_{(G)} A_{a\mu}A_b^{\mu}\phi_i + \frac{1}{4} \Lambda^{abij}_{(G)}
A_{a\mu}A_{b}^\mu\phi_i\phi_j \;,
\end{align}
where 
\begin{align}
\Lambda &= V^{(0)}(\omega_i) = L^i \omega_i + \frac{1}{2!} L^{ij}\omega_i \omega_j + 
\frac{1}{3!} L^{ijk} \omega_i \omega_j \omega_k + \frac{1}{4!} L^{ijkl} \omega_i 
\omega_j \omega_k \omega_l \label{Eq:Tree-Level} \\
\Lambda_{(S)}^i &= L^i + L^{ij} \omega_j + \frac{1}{2} L^{ijk}\omega_j \omega_k 
+ \frac{1}{6} L^{ijkl}\omega_j\omega_k\omega_l \\
\Lambda_{(S)}^{ij} &= L^{ij} + L^{ijk}\omega_k + \frac{1}{2}
L^{ijkl}\omega_k\omega_l  \label{eq:scalarten}\\
\Lambda_{(S)}^{ijk} &= L^{ijk}+L^{ijkl}\omega_l \\
\Lambda_{(S)}^{ijkl} &= L^{ijkl} \\
\Lambda_{(G)}^{ab} &= \frac{1}{2} G^{abij}\omega_i\omega_j 
  \label{eq:gaugeten} \\
\Lambda_{(G)}^{abi} &= G^{abij}\omega_j \\
\Lambda_{(G)}^{abij} &= G^{abij} \label{eq:gabijdef}\\
\Lambda_{(F)}^{IJ} &= M^{\ast IL} M_{L}^{\; J} = Y^{\ast  ILk}Y_L^{\; Jm} \omega_k\omega_m \;, \quad
\mbox{with} \label{eq:fermten}\\
M^{IJ} &= Y^{IJk}\omega_k \,.
\end{align}
Using this notation\footnote{For further details, we refer to
  \cite{Camargo-Molina:2016moz}.} one only needs to provide
$\omega_i,L^i,L^{ij},L^{ijk},L^{ijkl},G^{abij}$ and $Y^{IJk}$ to the program. 
 
\subsection{The Coleman-Weinberg Potential}
The temperature-independent one-loop corrected effective potential
in the Landau gauge is given by the Coleman-Weinberg
\cite{ColemanWeinberg} contribution as  
\begin{align}
V^{\text{CW}}(\omega) &= \frac{\varepsilon}{4} \summe{X={S,G,F}}{}
\left(-1\right)^{2s_X} \left(1+2s_X\right) \Tr\left[
\left(\Lambda^{xy}_{(X)}\right)^2 \left( \log\left(
\frac{1}{\mu^2} \Lambda^{xy}_{(X)} \right) - k_X
\right) \right] \label{Eq:CWPot} \;,
\end{align}
where $s_X$ denotes the spin of the particle described by the field $X$ and 
\begin{align}
\varepsilon &\equiv \frac{1}{\left(4\pi\right)^2} \,.
\end{align}
The indices $xy$ relate to the scalar indices $ij$, the gauge indices
$ab$ and the fermion indices $IJ$ for $X=S,G$ and $F$, respectively.
Note that the sum over $X$ has to be performed over all degrees of
freedom including the color degrees of freedom for the quarks. 
The scalar tensor $\Lambda_{(S)}^{ij}$, the gauge tensor
$\Lambda_{(G)}^{ab}$ and the fermion tensor $\Lambda_{(F)}^{IJ}$ are
given by Eq.~(\ref{eq:scalarten}), Eq.~(\ref{eq:gaugeten}) and
Eq.~(\ref{eq:fermten}), respectively. The potential is renormalized in
the $\overline{\mbox{MS}}$ scheme, {\it i.e.}~the default values for
the renormalization constants are
\beq
k_X = \left\{ \begin{array}{ll} \frac{5}{6} \;, & \quad \mbox{for
                                                  gauge bosons}
\\[0.1cm] \frac{3}{2} \;, & \quad \mbox{otherwise}
\end{array} \right. 
\eeq
In the program they are set in the file {\tt
  ClassPotentialOrigin.h} and named {\it C\_CWcbFermion},
{\it C\_CWcbGB} and {\it C\_CWcbHiggs} for the fermions, gauge bosons
and scalars, respectively. The renormalization scale $\mu$ is by default set to
  the VEV at $T=0$, $\mu = v(T=0) \approx 246.22$~GeV.

\subsection{The Counterterm Potential \label{sec:counterpot}}
The masses and mixing angles of the various involved particles are
derived from the loop-corrected potential and differ from the values
extracted from the tree-level potential. The tests for the
compatibility of the investigated model with the experimental
constraints have to implement these corrections. For an efficient scan
over the - often large - parameter space of the models it is
therefore more convenient to directly use loop-corrected masses and
angles as input. This is achieved by modifying the
$\overline{\mbox{MS}}$ renormalization of the Coleman-Weinberg
potential and applying the 
renormalization prescription by which the one-loop masses and mixing
angles are enforced to be equal to their values at tree-level. In practice,
we add the counterterm potential $V_{\text{CT}}$ implementing the
corresponding renormalization conditions. After
replacing the bare parameters $p^{(0)}$ of the tree-level
potential $V^{(0)}$ by the renormalized ones, $p$, and the
counterterms $\delta p$, it is given by 
\begin{align}
V^{\text{CT}} &= \sum_{i=1}^{n_p} \frac{\partial V^{(0)}}{\partial p_i} \delta p_i + 
\sum_{k=1}^{n_v} \delta T_k \left(\phi_k + \omega_k \right) \label{Eq:CTPot} \,,
\end{align}
where $n_p$ is the number of parameters of the potential. The $\delta
T_k$ denote the counterterms of the tadpoles $T_k$ obtained from the
minimum conditions of the potential for the $n_v$ directions in field space
in which we allow for the development of a non-zero vacuum expectation
value. Note, that $n_v \le n_{\text{Higgs}}$. In Sec.~\ref{sec:implmodels},
we give some explicit examples for counterterm potentials. The 
explicit forms of the finite counterterms are obtained from the renormalization 
conditions. Applying our renormalization prescription to the one-loop
contribution of the effective potential at $T=0$, 
{\it i.e.}~to $V^{\text{CW}}+V^{\text{CT}}$, yields the equations
($i,j=1,\dots ,n_v$)
\begin{align}
0 &= \left.\partial_{\phi_i} \left( V^{\text{CW}} + V^{\text{CT}}
\right)\right|_{\omega=\omega_{\text{tree}}} \label{Eq:Renorm1}
  \\ 
0 &= \left.\partial_{\phi_i} \partial_{\phi_j} \left( V^{\text{CW}} + V^{\text{CT}} \right)
\right|_{\omega =\omega_{\text{tree}}} \label{Eq:Renorm2} \,,
\end{align}
where $\omega_{\text{tree}}$ is the minimum of the tree-level
potential and $\omega$ stands generically for the
$n_v$ values $\omega_i$. The solution of the renormalization
conditions Eqs.~(\ref{Eq:Renorm1}) and 
(\ref{Eq:Renorm2}) requires the first and second derivatives of the
Coleman-Weinberg potential. The corresponding formulae have been derived in
\cite{Camargo-Molina:2016moz} and have been implemented in the
code. When a new model is added they can be obtained by calling
the functions {\tt WeinbergFirstDerivative} and {\tt WeinbergSecondDerivative}. 
If no shifts to the finite parts are needed, {\it
    i.e.}~if the $\overline{\mbox{MS}}$ scheme is applied, the
program will treat the finite parts of the counterterms as zero in the
new class corresponding to the new model.  

\subsection{The Thermal Corrections}
The temperature dependent potential $V^{(T)}$ is given by
\cite{Dolan:1973qd,Quiros:1999jp} 
\begin{align}
V^T (\omega,T) &= \summe{X={S,G,F}}{} (-1)^{2 s_X}
(1 + 2 s_X)\frac{T^4}{2\pi^2} J_{\pm}\left(\Lambda^{xy}_{(X)}/T^2
\right) \;, \label{Eq:TempPot}
\end{align}
with the functions $J_-$ for bosons and $J_+$ for fermions, respectively, reading
\begin{align}
J_{\pm}\left(\Lambda_{(X)}^{xy}/T^2\right) &= \Tr\left[ \bint{0}{\infty} \,\mathrm{dk}\,  
k^2 \log\left[ 1 \pm \exp\left( -\sqrt{k^2 + \Lambda^{xy}_{(X)}/T^2}\right) \right] \right]
\,. 
\end{align}
Furthermore we have to calculate the daisy corrections \cite{Carrington:1991hz}
$\Pi_{(S)}^{ij}$ and $\Pi_{(G)}^{ab}$ to the masses of the scalars and
gauge bosons, respectively. They are given by  
\begin{align}
\Pi_{(S)}^{ij} =& \frac{T^2}{12} \left[ \left(-1\right)^{2s_S} \left(1+2s_S\right) 
\summe{k=1}{n_{\text{Higgs}}} L^{ijkk} + \left(-1\right)^{2s_G} \left(1+2s_G\right) 
\summe{a=1}{n_{\text{gauge}}} G^{aaij}  \right. \notag \\
& \left. + \left(-1\right)^{2s_F} \left(1+2s_F\right) \frac{1}{2} 
\summe{I,J=1}{n_{\text{fermion}}} \left(Y^{\ast
  IJj}Y_{IJ}^j + Y^{\ast IJi}Y_{IJ}^{j} \right) \right]  \label{eq:pis}\\
\Pi_{(G)}^{ab} =& T^2 \frac{2}{3} \left(\frac{\tilde{n}_H}{8} + 5 \right) \frac{1}{\tilde{n}_H}
\summe{m=1}{n_{\text{Higgs}}} \Lambda^{aamm}_{(G)} 
\delta_{ab} \,, \label{eq:pigauge}
\end{align}
where only the longitudinal modes of the gauge bosons get the daisy
corrections and $\tilde{n}_H \le n_{\text{Higgs}}$ is the number of
Higgs fields coupling to the gauge bosons. 
The tensors $L^{ijkk}$, $Y^{IJi}$ and $G^{aaij}$ have
  been introduced in Eq.~(\ref{Eq:LS_Classical}),
  Eq.~(\ref{Eq:LF_Classical}) and Eq.~(\ref{Eq:LG_Classical}), respectively. The
tensor $\Lambda^{aamm}_{(G)}$ has been defined in Eq.~(\ref{eq:gabijdef}).
There are two methods to evaluate these corrections.
\begin{itemize}
\item According to the Arnold-Espinosa method \cite{Arnold:1992rz}
  one makes the replacement  
\begin{align}
V^T (\omega,T) &\to V^T (\omega,T) + V_{\text{daisy}}(\omega,T) \label{eq:AE1}\,,\\
V_{\text{daisy}}(\omega,T) &= -\frac{T}{12\pi} \left[ \summe{i=1}{n_{\text{Higgs}}}
\left((\overline{m}^2_i)^{3/2}  - (m_i^2)^{3/2}\right) +\summe{a=1}{n_{\text{gauge}}} 
\left((\overline{m}^2_a)^{3/2} - (m_a^2)^{3/2} 
\right) \right]  \label{eq:AE2} 
\end{align}
where
$m_i^2,\overline{m}_i^2, m_a^2, \overline{m}_a^2$ are the eigenvalues
of $\Lambda_{(S)}^{ij}, \Lambda_{(S)}^{ij} + \Pi^{ij}_{(S)} ,
\Lambda_{(G)}^{ab}, \Lambda_{(G)}^{ab} +
\Pi^{ab}_{(G)}$. Remark, that only the longitudinal modes of the gauge
bosons get the thermal corrections $\Pi^{ab}_{(G)}$.
Note also that $V^T (\omega,T)$ only depends
  on masses excluding the thermal corrections.
\item In the Parwani method \cite{Parwani:1991gq}, on the other hand, one replaces 
\begin{align}
\Lambda_{(S)}^{ij} &\to \Lambda_{(S)}^{ij} + \Pi^{ij}_{(S)} \label{eq:Parwani1}
\end{align}
in \eqref{Eq:CWPot} and \eqref{Eq:TempPot} and also
\begin{align}
\Lambda_{(G)}^{ab} &\to \Lambda_{(G)}^{ab} + \Pi^{ab}_{(G)} \label{eq:Parwani2}
\end{align}
for the longitudinal modes. The Debye corrected masses are hence also
used in $V^{\text{CW}}$.
\end{itemize}
	
\subsection{Treatment of $J_\pm$ \label{Sec:Jpm}}
The numerical evaluation of\footnote{For a recent {\tt C++} library
  for the computation of these functions, see \cite{Fowlie:2018eiu}.} 
\begin{align}
J_\pm(x^2) &= \bint{0}{\infty} \mathrm{dk}\, k^2 \log\left[ 1 \pm
\exp\left(-\sqrt{k^2+x^2}\right)\right] 
\end{align}
is very time consuming and
therefore we use the series expansions in small $x^2 = m^2/T^2$, 
\begin{align}
J_{+,s}(x^2,n) =& - \frac{7\pi^4}{360} + \frac{\pi^2}{24} x^2 +\frac{1}{32} x^4
\left(\log x^2 - c_+\right) \notag
\\ &- \pi^2 x^2 \summe{l=2}{n} \left( - \frac{1}{4\pi^2} x^2\right)^l
\frac{\left(2l-3\right)!!\zeta\left(2l-1\right)}{\left(2l\right)!!\left(l+1\right)}
 \left(2^{2l-1}-1\right) \\ 
J_{-,s}(x^2,n) =& - \frac{\pi^4}{45} + \frac{\pi^2}{12} x^2 - \frac{\pi}{6} \left(x^2
\right)^{3/2} - \frac{1}{32} x^4 \left(\log x^2 - c_-\right) \notag \\ &+
\pi^2x^2 \summe{l=2}{n} \left(-\frac{1}{4\pi^2} x^2\right)^l\frac{\left(2l-3\right)!!
\zeta\left(2l-1\right)}{\left(2l\right)!!\left(l+1\right)} \,, 
\end{align}
with
\begin{align}
c_+ &= \frac{3}{2} + 2\log \pi - 2\gamma_E \\
c_- &= c_+ + 2\log 4 \,,
\end{align}
where $\gamma_E$ denotes the Euler-Mascheroni constant, $\zeta(x)$ the
Riemann $\zeta$-function and $(x)!!$ the double factorial. For large $x^2$ we use
\begin{align}
J_{\pm,l}(x^2,n) &= -\exp\left(-\left(x^2\right)^{1/2}\right)\left(\frac{\pi}{2} \left(x^2
\right)^{3/2}\right)^{1/2} \summe{l=0}{n} \frac{1}{2^l l!}
\frac{\Gamma\left(5/2+l\right)}{\Gamma\left(5/2-l\right)} \left(x^2\right)^{-l/2} \,.
\end{align}
With 
\begin{align}
x_+^2 &= 2.2161\,, \qquad \delta_+ = -0.015603 \,, \\
x_-^2 &= 9.4692\,, \qquad \delta_- = 0.0063109 \,,
\end{align}
we then calculate $J_\pm$ as 
\begin{align}
J_+(x^2) &= \begin{cases} - J_{\pm,l}(x^2,3) & x^2 \geq x_+^2 \\
  -\left(J_{+,s}(x^2,4) + \delta_+ \right)  & x < x_+^2 \end{cases} \\
J_-(x^2) &= \begin{cases} J_{\pm,l}(x^2,3) & x^2 \geq x_-^2 \\  J_{-,s}(x^2,3) + \delta_-   
& x^2 < x_-^2 \end{cases} \;.
\end{align}
The shifts $\delta_\pm$ arise because we choose the intersection point of
$J_s$ and $J_l$ to be such that the derivatives are continuous, and
these shifts then enforce the functions themselves to be continuous.  
In the course of the scan over the parameter space it can happen that
the bosonic masses become negative, so that $J_-(x^2)$ will be called
for $x^2 < 0$.
In this case, only the real part of the function is taken \cite{Weinberg:1987vp}. In
practice, the integral is evaluated numerically from
  $x^2=0$ down to $x^2 = -3000$ in steps of 1. 
In the minimization procedure the result obtained from the linear
interpolation between these points is then used. \s
\begin{figure}[t]
\centering
\subfigure[The integral $J_-$ for the bosons and $x^2\geq
0$.]{\includegraphics[scale=0.5]{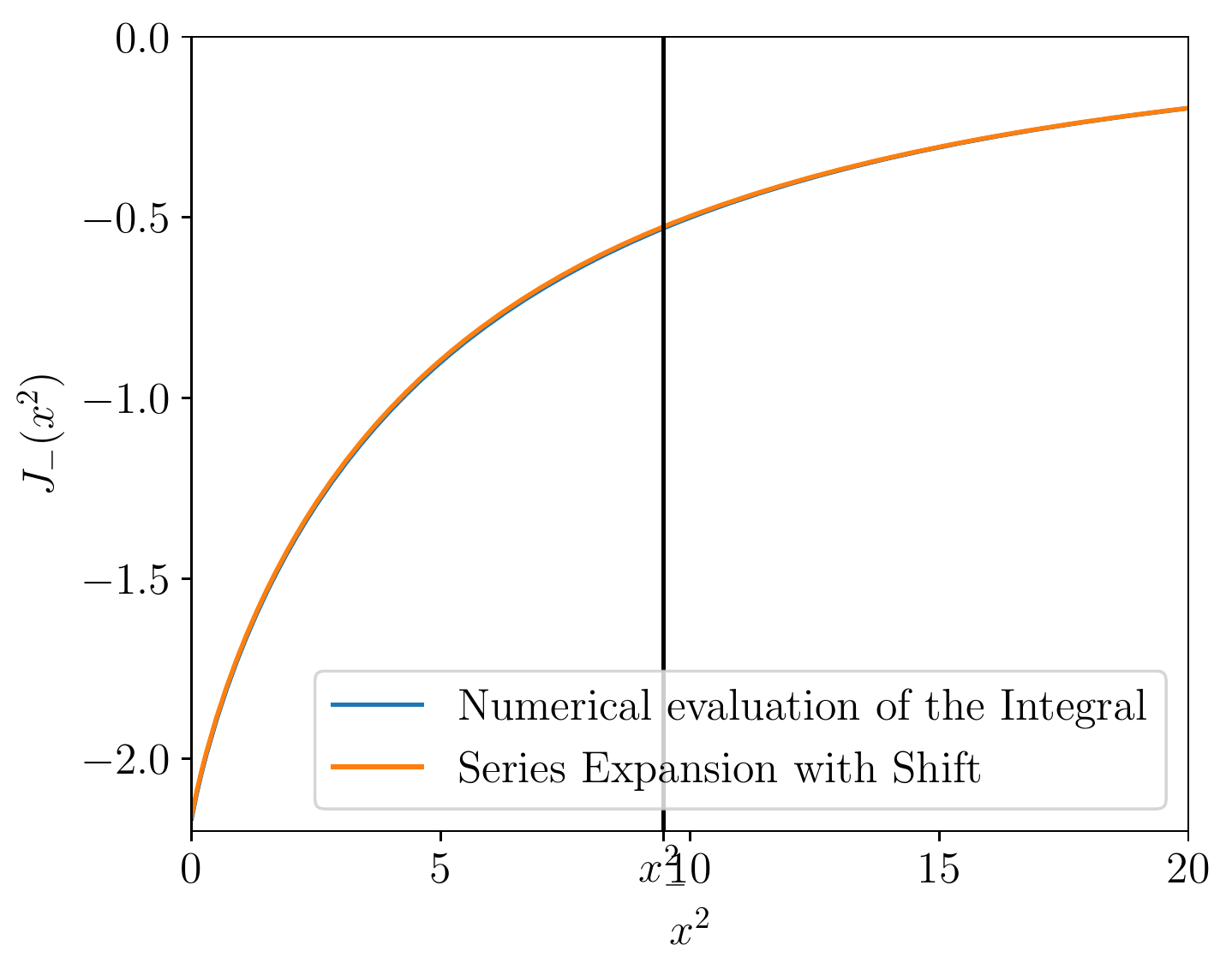}} \qquad
\subfigure[The integral $J_+$ for the
fermions.]{\includegraphics[scale=0.5]{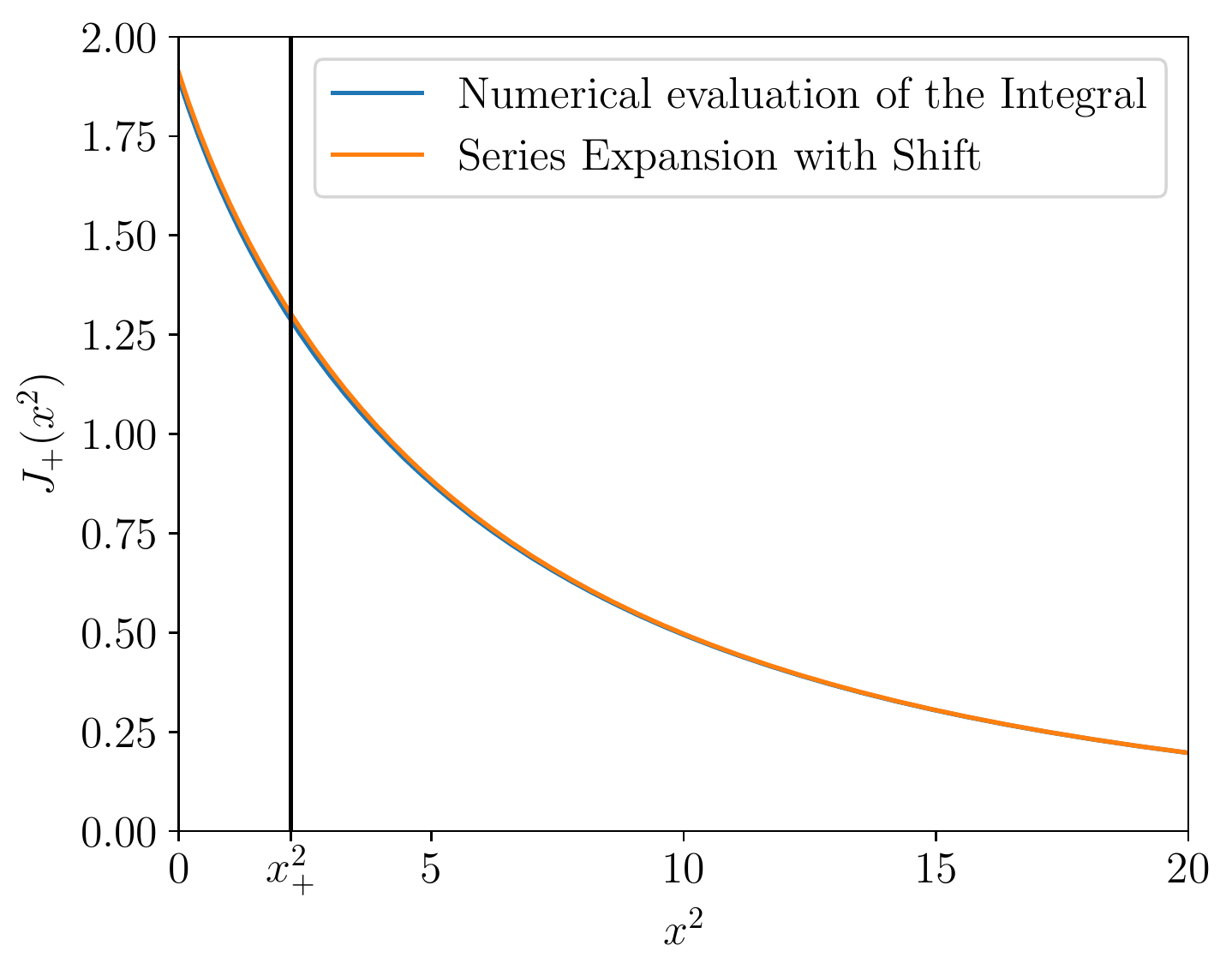}} \\
\subfigure[The relative difference in $J_-$ for the series expansion
($S$) and the numerical
evaluation ($I$).]{\includegraphics[scale=0.5]{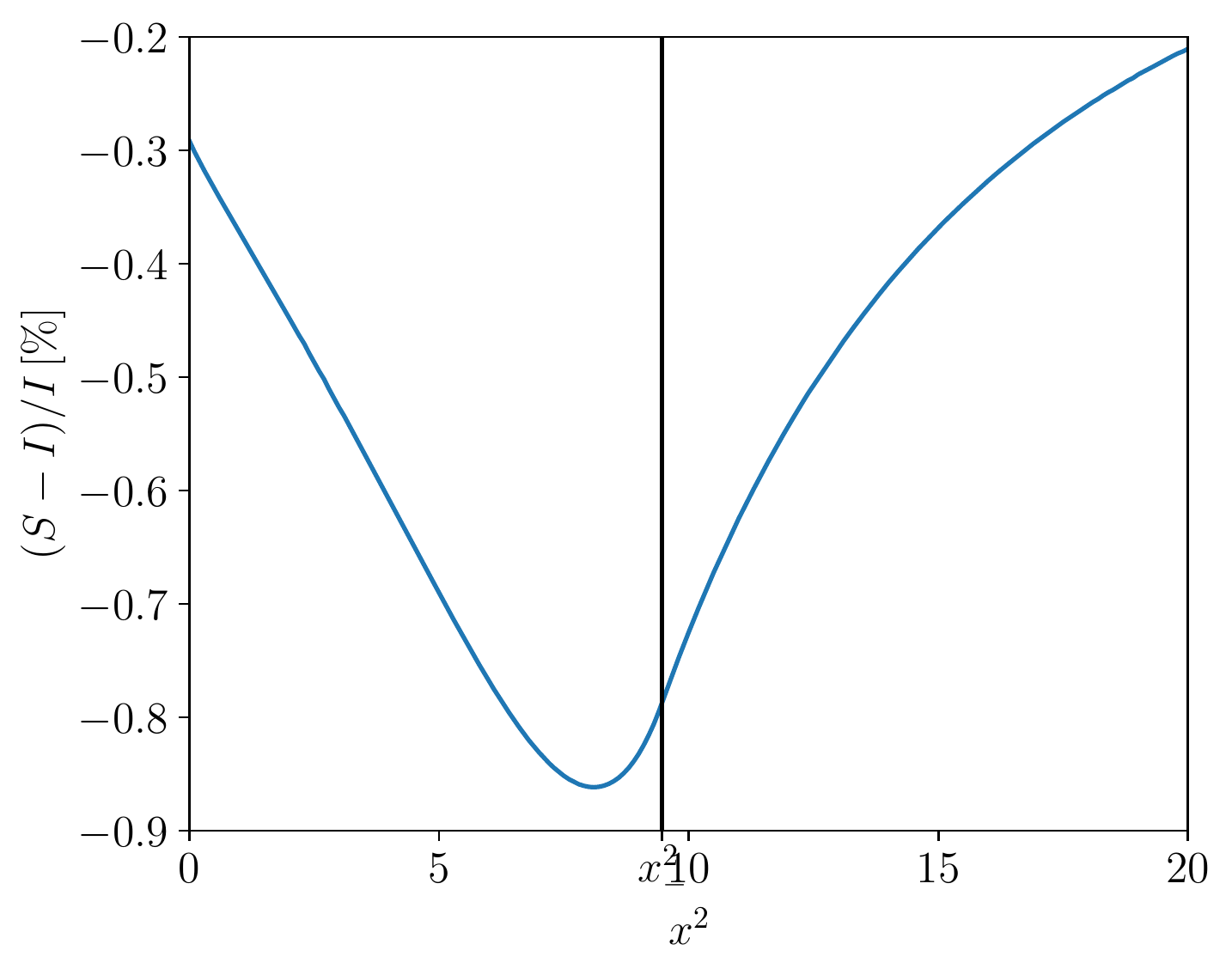}}
\qquad 
\subfigure[The relative difference in
$J_+$.]{\includegraphics[scale=0.5]{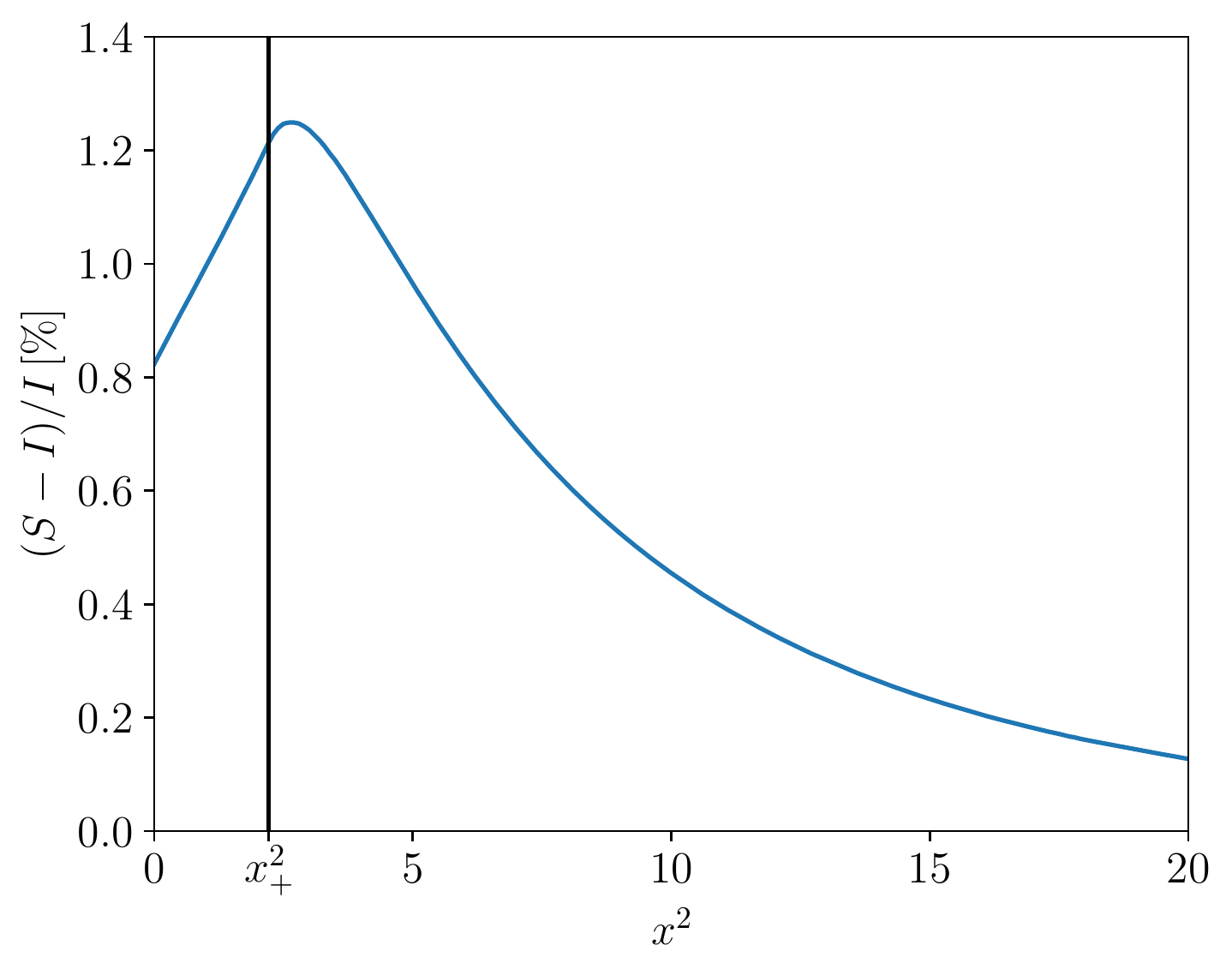}} 
\caption{
Comparison between the numerical integration and the series
  expansion of $J_\pm(x^2)$ around $x_\pm^2$. \label{fig:CompareJpm}} 
\end{figure}
	
In Fig.~\ref{fig:CompareJpm} (upper) we show the series expansion
around the transition point $x_-^2$ ($x_+^2$) for $J_-$ ($J_+$) on the
left (right) side compared to the numerical evaluation of $J_-$
($J_+$). The plots in the lower row show the relative difference
between the series expansion $S$ and the numerical evaluation $I$,
$(S-I)/I$ in per cent. As can be inferred from
Fig.~\ref{fig:CompareJpm} (c), it does not exceed  
1\% in case of bosons, and for fermions it exceeds $1\%$ only around
the transition point, {\it cf.}~Fig.~\ref{fig:CompareJpm} (d). 

\subsection{The Minimization of the Effective
  Potential \label{sec:minimize}}
For the EWPT to be considered of strong first order, the ratio of the VEV $v_c$ at
the critical temperature $T_c$ has to fulfill $v_c/T_c > 1$. The
value $v$ of the VEV at a given temperature $T$ is obtained as
\beq
v(T) = \left(\sum_{k=1}^{\tilde{n}_H} \bar{\omega}_k^2
\right)^{\frac{1}{2}} \;.
\label{eq:ewvev}
\eeq
Here, $\tilde{n}_H$ means that the sum is performed over all 
directions in field space in which we allow for the development of a
non-zero {\it electroweak} VEV, {\it i.e.}~the VEV for fields that couple to the EW gauge
bosons. We hence do not include here the VEV that a gauge
singlet field (as it appears for example in the Next-to-2HDM)
develops. The $\bar{\omega}_k$ denote the field configurations that minimize
the loop-corrected effective potential. Therefore,
Eq.~(\ref{eq:ewvev}) is the VEV that coincides at $T=0$ with $v=246.22$~GeV.
The critical temperature $T_c$
is the temperature where two degenerate minima of the potential
exist. In order to determine $T_c$ the effective potential including
the counterterm potential, {\it cf}.~Eq.~(\ref{eq:fielconfig}), is
minimized numerically at a given temperature $T$. In case of a first
order EWPT the VEV jumps from $v= v_c$ at the temperature $T_c$ to
$v=0$ at $T > T_c$. For the minimization we use the algorithm {\tt
  CMAES} as implemented in {\tt libcmaes} \cite{CMAES}, which 
finds the global minimum of a given function. As termination criterion
we require the relative tolerance of the value of the effective
potential between two iterations to be below $10^{-5}$.
For the determination of $T_c$ we employ a bisection method in the
interval $T \in [0,300]  \,\mathrm{GeV}$ until the interval containing
$T_c$ is smaller than $10^{-2}$ GeV. The temperature $T_c$ is then set
to the lower bound of the final interval. We exclude parameter points
for which the individual VEVs obtained from the next-to-leading order
(NLO) potential at $T=0$ deviate by more than 1~GeV from their input
values as well as parameter points where no PT is found for $T\le 300$~GeV. 

\section{Implemented Models \label{sec:implmodels}}
In this section we provide the tree-level potentials as well as the
counterterms for the already implemented models. For all implemented
models the code expects an input file that presents the input
parameters in the same way the program code {\tt ScannerS}
\cite{Coimbra:2013qq,scanners2}  writes them into its default output files. 
{\tt ScannerS} is a program that allows to perform extensive scans in
the parameter space of multi-Higgs models and checks for compatibility
with theoretical and experimental constraints. The viable parameter
points can then be fed in our program to investigate the compatibility
with a strong first order EWPT {\it e.g.}. So far, we have applied our code
for such an analysis in the CP-conserving or real 2HDM (R2HDM)
\cite{Basler:2016obg}, the CP-violating or complex 2HDM (C2HDM)
\cite{Basler:2017uxn} and the Next-to-2HDM (N2HDM) 
\cite{jonasmueller}. 

 
\subsection{The CP-Conserving 2HDM}
The tree-level Higgs potential of the CP-conserving 2HDM
\cite{Lee:1973iz,Branco:2011iw} with a softly broken $\mathbb{Z}_2$
symmetry, under which the two $SU(2)_L$ Higgs doublets
$\Phi_1$ and $\Phi_2$,
\beq
\Phi_1 = \begin{pmatrix} \phi_1^+ \\ \phi_1^0 \end{pmatrix}
\quad \mbox{and} \quad
\Phi_2 = \begin{pmatrix} \phi_2^+ \\ \phi_2^0 \end{pmatrix} \;,
\eeq
transform as $\Phi_1 \to \Phi_1$, $\Phi_2 \to - \Phi_2$, reads
\begin{eqnarray}
V^{(0)} &=& m_{11}^2 \Phi_1^\dagger \Phi_1 + m_{22}^2 \Phi_2^\dagger
 \Phi_2 - m_{12}^2\left(\Phi_1^\dagger\Phi_2 + h.c. \right) + \frac{\lambda_1}{2}
\left(\Phi_1^\dagger\Phi_1\right)^ 2 + \frac{\lambda_2}{2}
\left(\Phi_2^\dagger\Phi_2\right)^2 \notag \\ 
&&+ \lambda_3\left(\Phi_1^\dagger\Phi_1\right)\left(\Phi_2^\dagger\Phi_2\right) +
\lambda_4\left(\Phi_1^\dagger\Phi_2\right)\left(\Phi_2^\dagger\Phi_1\right) +
\frac{\lambda_5}{2} \left[\left(\Phi_1^\dagger\Phi_2\right)^2 + h.c. \right] \;.
\end{eqnarray}
The mass parameters $m_{11}^2$, $m_{22}^2$ and $m_{12}^2$ as well as the 
quartic couplings $\lambda_1 ... \lambda_5$ are real. The parameters
$m_{12}^2$ and $\lambda_5$ can be complex in the CP-violating
2HDM. After EWSB the two Higgs doublets acquire VEVs $\bar{\omega}
\in \mathbb{R}$
about which the Higgs fields can be
expanded in terms of the charged field components $\rho_i$ and
$\eta_i$ and the neutral CP-even and CP-odd fields $\zeta_i$ and
$\psi_i$ ($i=1,2$),
\begin{eqnarray}
\Phi_1 &=& \frac{1}{\sqrt{2}} \begin{pmatrix} \rho_1+\ii\eta_1 \\
  \zeta_1 + \bar{\omega}_1 + \ii \psi_1 \end{pmatrix} \\
\Phi_2 &=& \frac{1}{\sqrt{2}} \begin{pmatrix} \rho_2 + \bar{\omega}_{\text{CB}} +\ii\eta_2 
\\ \zeta_2+\bar{\omega}_2 +
\ii\left(\psi_2+\bar{\omega}_{\text{CP}}\right) \end{pmatrix}  \;.
\end{eqnarray}
In order to be as general as possible we also allow for
CP-violating ($\bar{\omega}_{\text{CP}}$) and charge-breaking
($\bar{\omega}_{\text{CB}}$) VEVs although the latter obviously is unphysical.
Note that without loss of generality we have rotated the complex part
of the VEVs to the second doublet exclusively. We denote the VEVs of
our present vacuum by ($i=1,2,\mbox{CP},\mbox{CB}$)
\beq
v_i \equiv \bar{\omega}_i|_{T=0} \;,
\eeq
with
\beq
v_{\text{CP}} = v_{\text{CB}} = 0 \;,
\eeq
while the remaining two VEVs are related to the SM VEV $v \approx
246.22$~GeV through
\beq
v_1^2 + v_2^2 = v^2 \;.
\eeq
By introducing the angle $\beta$ as
\beq
\tan \beta = \frac{v_2}{v_1} \;
\eeq
we have 
\beq
v_1 = v \cos\beta \quad \mbox{and} \quad v_2 = v \sin \beta \;.
\eeq
The mixing angle $\beta$ is the rotation angle from the gauge to the
mass eigenstates in the charged and in the CP-odd sector,
respectively, while we call the mixing angle in the CP-even sector
$\alpha$,
\beq
\left( \begin{array}{c} G^\pm \\ H^\pm \end{array} \right) = R(\beta) 
\left( \begin{array}{c} \phi_1^\pm \\ \phi_2^\pm \end{array} \right) 
\;, \quad
\left( \begin{array}{c} G^0 \\ A \end{array} \right) = R(\beta) 
\left( \begin{array}{c} \psi_1 \\ \psi_2 \end{array} \right)  \;,
\quad
\left( \begin{array}{c} H \\ h \end{array} \right) = R(\alpha) 
\left( \begin{array}{c} \zeta_1 \\ \zeta_2 \end{array} \right)  \;,
\eeq
with
\beq
R(x) = \left( \begin{array}{cc} \cos x & \sin x \\ - \sin x & \cos x \end{array}\right) \;.
\eeq
We have five physical mass eigenstates, the light and heavy CP-even
Higgs bosons $h$ and $H$, the pseudoscalar $A$ and a charged Higgs
pair $H^\pm$, while $G^0$ and $G^\pm$ represent the neutral and
charged massless Goldstone bosons. The counterterm potential is given as
\begin{align}
V^{\text{CT}} =& \delta m_{11}^2 \Phi_1^\dagger \Phi_1 + \delta m_{22}^2
\Phi_2^\dagger \Phi_2 - \delta  m_{12}^2\left(\Phi_1^\dagger\Phi_2 + \Phi_2^\dagger
\Phi_1\right) + \frac{\delta \lambda_1}{2}\left(\Phi_1^\dagger\Phi_1\right)^ 2 + 
\frac{\delta \lambda_2}{2} \left(\Phi_2^\dagger\Phi_2\right)^2 \notag \\
&+ \delta\lambda_3\left(\Phi_1^\dagger\Phi_1\right)\left(\Phi_2^\dagger\Phi_2\right)
+ \delta \lambda_4\left(\Phi_1^\dagger\Phi_2\right)\left(\Phi_2^\dagger\Phi_1\right)
+ \frac{\delta \lambda_5}{2} \left[ \left(\Phi_1^\dagger\Phi_2\right)^2 +
\left(\Phi_2^\dagger\Phi_1\right)^2\right] \notag \\ 
&+ \delta T_1\left(\zeta_1 + \omega_1 \right) + \delta
  T_2\left(\zeta_2+\omega_2\right) + \delta
 T_{\text{CP}} \left( \psi_2+\omega_{\text{CP}}\right) + \delta T_{\text{CB}}
  \left(\rho_2 + \omega_{\text{CB}} \right) \,.  
\end{align}
In order to avoid flavour-changing neutral currents (FCNC) at tree level, the
$\mathbb{Z}_2$ symmetry can also be extended to the Yukawa
sector \cite{Glashow:1976nt,Paschos:1976ay}. With four possible
$\mathbb{Z}_2$ charge assignments there are four different types of
2HDMs as summarized in Table~\ref{tab:chargassign}. \s
\begin{table}
\begin{center}
\begin{tabular}{rccc} \toprule
& $u$-type & $d$-type & leptons \\ \midrule
Type I & $\Phi_2$ & $\Phi_2$ & $\Phi_2$ \\
Type II & $\Phi_2$ & $\Phi_1$ & $\Phi_1$ \\
Lepton-Specific & $\Phi_2$ & $\Phi_2$ & $\Phi_1$ \\
Flipped & $\Phi_2$ & $\Phi_1$ & $\Phi_2$ \\ \bottomrule
\end{tabular}
\caption{The four Yukawa types of the softly broken
  $\mathbb{Z}_2$-symmetric 2HDM. \label{tab:chargassign}} 
\end{center}
\end{table}
 
The on-shell renormalization that we apply leads to the conditions
\beq
\left.\partial_{\phi_i} V^{\text{CT}}\right|_{\phi =\langle \phi^c\rangle_{T=0}} &=&
-\left.\partial_{\phi_i} V^{\text{CW}}\right|_{\phi =\langle \phi^c \rangle_{T=0}} \\
\left.\partial_{\phi_i}\partial_{\phi_j} V^{\text{CT}}
\right|_{\phi=\langle \phi^c \rangle_{T=0}} &=& -\left.\partial_{\phi_i}
\partial_{\phi_j} V^{\text{CW}} \right|_{\phi=\langle\phi^c \rangle_{T=0}}  \,, 
\eeq
where
\beq
\phi_i \equiv \{ \rho_1, \eta_1, \rho_2, \eta_2, \zeta_1, \psi_1,
\zeta_2, \psi_2 \} \label{eq:vec2hdm}
\eeq
and $\langle \phi^c \rangle_{T=0}$ denotes the field configuration in
the minimum at $T=0$,
\beq
\langle \phi^c \rangle_{T=0} = (0,0,0,0,v_1,0,v_2,0) \;.
\eeq
These conditions yield the counterterms
\begin{align}
\delta m_{11}^2 &= \frac{1}{2} H^{\CW}_{\zeta_1,\zeta_1} +H^{\CW}_{\psi_1,\psi_1} - \frac{5}{2}
H^{\CW}_{\rho_1,\rho_1} + \frac{1}{2} \frac{v_2}{v_1}\left( H^{\CW}_{\zeta_1,\zeta_2} -
H^{\CW}_{\eta_1,\eta_2}  \right) + t v_2^2 \\
\delta m_{22}^2 &= \frac{1}{2} \left( H^{\CW}_{\zeta_2,\zeta_2} -3H^{\CW}_{\eta_2,\eta_2} \right) + 
\frac{1}{2} \frac{v_1}{v_2} \left( H^{\CW}_{\zeta_1,\zeta_2} -
H^{\CW}_{\eta_1,\eta_2}\right) + \frac{v_1^2}{v_2^2}\left( H^{\CW}_{\psi_1,\psi_1} -
H^{\CW}_{\rho_1,\rho_1}\right) + v_1^2 t \\
\delta m_{12}^2 &= H^{\CW}_{\eta_1,\eta_2} + \frac{v_1}{v_2} \left(
H^{\CW}_{\psi_1,\psi_1} - H^{\CW}_{\rho_1,\rho_1}\right) + v_1v_2 t \\
\delta \lambda_1 &= \frac{1}{v_1^2} \left[ 2 H^{\CW}_{\rho_1,\rho_1} -
H^{\CW}_{\zeta_1,\zeta_1} - H^{\CW}_{\psi_1,\psi_1}\right] - \frac{v_2^2}{v_1^2} t \\
\delta \lambda_2 &= \frac{1}{v_2^2} \left[ H^{\CW}_{\eta_2,\eta_2} -
H^{\CW}_{\zeta_2,\zeta_2}\right) +\frac{v_1^2}{v_2^4} \left( H^{\CW}_{\rho_1,\rho_1} -
H^{\CW}_{\psi_1,\psi_1}\right) - \frac{v_1^2}{v_2^2} t \\
\delta \lambda_3 &= \frac{1}{v_1v_2} \left( H^{\CW}_{\eta_1,\eta_2}-H^{\CW}_{\zeta_1,\zeta_2}
\right) + \frac{1}{v_2^2}\left(H^{\CW}_{\rho_1,\rho_1} -H^{\CW}_{\psi_1,\psi_1}\right) - t \\
\delta \lambda_4 &= t \\
\delta \lambda_5 &= \frac{2}{v_2^2}
\left(H^{\CW}_{\psi_1,\psi_1}-H^{\CW}_{\rho_1,\rho_1}  \right) + t \\
\delta T_1 &= H^{\CW}_{\eta_1,\eta_2} v_2 + H^{\CW}_{\rho_1,\rho_1} v_1 - N^{\CW}_{\zeta_1} \\
\delta T_2 &= H^{\CW}_{\eta_1,\eta_2} v_1 + H^{\CW}_{\eta_2,\eta_2} v_2 - N^{\CW}_{\zeta_2} \\
\delta T_{\text{CP}} &= \frac{v_1^2}{v_2} H^{\CW}_{\zeta_1,\psi_1} + H^{\CW}_{\zeta_1,\psi_2} v_1 - 
N^{\CW}_{\psi_2}   \\  
\delta T_{\text{CB}} &= -N^{\CW}_{\rho_2} \,,
\end{align}
where we used
\begin{align}
N^{\CW}_{\phi} &= \left.\partial_{\phi} V^{\CW}\right|_{\phi=\langle
                 \phi^c\rangle_{T=0}} \label{eq:nabbr} \\
H^{\CW}_{\phi_1,\phi_2} &=\left.\partial_{\phi_1}\partial_{\phi_2} V^{\CW}
\right|_{\phi_=\langle \phi^c \rangle_{T=0}} \,. \label{eq:habbr}
\end{align}
Having less renormalization constants than renormalization conditions,
the system of equations is overconstrained. Its consistent solution
is given by the following identities
\begin{align}
0 &= H^{\CW}_{\eta_1,\eta_1} - H^{\CW}_{\rho_1,\rho_1} \\
0 &= H^{\CW}_{\eta_1,\eta_2} - H^{\CW}_{\rho_1,\rho_2} \\
0 &= H^{\CW}_{\eta_2,\eta_2} - H^{\CW}_{\rho_2,\rho_2} \\
0 &= H^{\CW}_{\psi_1,\psi_2} - H^{\CW}_{\eta_1,\eta_2} + \frac{v_1}{v_2}
    \left( H^{\CW}_{\psi_1,\psi_1} - H^{\CW}_{\rho_1,\rho_1} \right) \\
0 &= H^{\CW}_{\psi_2,\psi_2} - H^{\CW}_{\eta_2,\eta_2} +\frac{v_1^2}{v_2^2} 
\left( H^{\CW}_{\rho_1,\rho_1} - H^{\CW}_{\psi_1,\psi_1}\right) \,,
\end{align}
leading to a one-dimensional solution space parametrized by the
parameter $t \in \mathbb{R}$. In the code $t$ is chosen such that
\begin{align}
	\delta \lambda_4 &= 0 \,.
\end{align}
Note that the renormalization constants $\delta T_{\text{CP}}$
  and $\delta T_{\text{CB}}$ always turn out to be zero as we do not have
CP violation\footnote{We set the CKM matrix to unity and
  hence do not have explicit CP violation in the model.} nor 
charge breaking.  \s

For the eight parameters of the Higgs potential we can either
choose a more 'physics' inspired set involving the masses of the physical
Higgs bosons or a pure 'parametric' input set. The code requires the
'parametric' input based on $\lambda_{1...5}, m_{12}^2$ and
$\tan\beta$\footnote{The eighth parameter is the SM VEV $v$ that is
  hard-coded in the program.}, which has to be given in the order 
\beq
\mbox{type} \;, \; \lambda_1 \;, \; \lambda_2 \;, \; \lambda_3 \;, \;
\lambda_4 \;, \; \lambda_5 \;, \; m_{12}^2 \;, \; \tan\beta \;.   
\eeq
%
The user furthermore has to specify through $\mbox{type} = 1,...,4$
the type of the 2HDM to be applied, 
as given in Table~\ref{tab:chargassign} where $\mbox{type} = 1,...,4$
corresponds to type I, type II, lepton-specific and flipped. 
Note that the minimum conditions of the potential lead to the following
relations among the parameters
\begin{align}
m_{11}^2 &= m_{12}^2 \frac{v_2}{v_1} - \frac{v_1^2}{2} \lambda_1^2 -
\frac{v_2^2}{2} \left(\lambda_3+\lambda_4+\lambda_5\right) \\
m_{22}^2 &= m_{12}^2 \frac{v_1}{v_2} - \frac{v_2^2}{2} \lambda_2^2 -
\frac{v_1^2}{2} \left(\lambda_3+\lambda_4+ \lambda_5\right) \,.
\end{align}

\subsection{The CP-violating 2HDM \label{sec:C2HDM}}
Incorporating the softly broken ${\mathbb Z}_2$ symmetry to avoid
FCNC at tree-level (implying the same four different types of 2HDM as
in the CP-conserving case, {\it cf.}~Table~\ref{tab:chargassign}), the
tree-level Higgs potential of the C2HDM 
\cite{Ginzburg:2002wt}\footnote{For recent phenomenological analyses,
  see \cite{Fontes:2014xva,c2hdmpheno,phenocomp}.} reads 
\begin{align}
\begin{split}
V^{(0)} &= m_{11}^2 \Phi_1^\dagger \Phi_1 + m_{22}^2
\Phi_2^\dagger \Phi_2 - \left[m_{12}^2 \Phi_1^\dagger \Phi_2 +
  \mathrm{h.c.} \right] + \frac{1}{2} \lambda_1 ( \Phi_1^\dagger
\Phi_1)^2 +\frac{1}{2} \lambda_2 (\Phi_2^\dagger \Phi_2)^2 \\
&\quad + \lambda_3 (\Phi_1^\dagger \Phi_1)(\Phi_2^\dagger\Phi_2) +
\lambda_4 (\Phi_1^\dagger \Phi_2)(\Phi_2^\dagger \Phi_1)
+ \left[ \frac{1}{2} \lambda_5 (\Phi_1^\dagger\Phi_2)^2  +
  \mathrm{h.c.} \right] \; .
\end{split}\label{eq:treec2hdmpot}
\end{align}
In contrast to the CP-conserving 2HDM, the two parameters $m_{12}^2$ and $\lambda_5$
can now be complex. If $\mbox{arg}(m_{12}^2)=\mbox{arg}(\lambda_5)$ 
the complex phases of these two parameters can be absorbed by a basis
transformation. If additionally the VEVs of the doublets are assumed
to be real, we have the real 2HDM. Otherwise, we are in the C2HDM. In
the following, we will adopt the conventions of \cite{Fontes:2014xva}.
After EWSB the two Higgs doublets develop VEVs and allowing for the
most general vacuum configuration, the expansion about the minimum
reads
\begin{eqnarray}
\Phi_1 &=& \frac{1}{\sqrt{2}} \begin{pmatrix} \rho_1+\ii\eta_1 \\
  \zeta_1 + \bar{\omega}_1 + \ii \psi_1 \end{pmatrix} \\
\Phi_2 &=& \frac{1}{\sqrt{2}} \begin{pmatrix} \rho_2 + \bar{\omega}_{\text{CB}} +\ii\eta_2 
\\ \zeta_2+\bar{\omega}_2 +
\ii\left(\psi_2+\bar{\omega}_{\text{CP}}\right) \end{pmatrix}  \;.
\end{eqnarray}
After introducing
\beq
\zeta_3 = - \psi_1 \sin\beta + \psi_2 \cos \beta
\eeq
the neutral mass eigenstates $H_i$ ($i=1,2,3$) are obtained from the
C2HDM basis $\zeta_1,$ $\zeta_2$ and $\zeta_3$ through the rotation
\beq
\left( \begin{array}{c} H_1 \\ H_2 \\ H_3 \end{array} \right)  =
R \left( \begin{array}{c} \zeta_1 \\ \zeta_2 \\ \zeta_3 \end{array} \right)  \;.
\eeq
The mass matrix $R$ can be parametrized in terms of three mixing
angles $\alpha_i$ ($i=1,2,3$) with $-\pi/2 \le \alpha_i < \pi/2$ as
\beq
R = \left( \begin{array}{ccc} c_1 c_2 & s_1 c_2 & s_2 \\ - (c_1 s_2
s_3 + s_1 c_3) & c_1 c_3 - s_1 s_2 s_3 & c_2 s_3 \\ -c_1 s_2 c_3 +
s_1 s_3 & - (c_1 s_3 + s_1 s_2 c_3) & c_2 c_3 \end{array} \right) \;.
\eeq
All neutral Higgs bosons mix and have no definite CP quantum
number. The masses are obtained from the diagonalization of the mass
matrix, derived from the Higgs potential, and the conventions are such
that $m_{H_1} \le m_{H_2} \le m_{H_3}$. 
The charged sector does not change with respect to
the CP-conserving 2HDM, and the mixing angle diagonalizing the charged
mixing matrix is given by $\beta$. \s

The counterterm potential reads
\begin{align}
V^{\text{CT}} =& \delta m_{11}^2 \Phi_1^\dagger \Phi_1 + \delta m_{22}^2
\Phi_2^\dagger \Phi_2 - \left(\delta \mbox{Re}(m_{12}^2) + \ii
\delta \mbox{Im}(m_{12}^2)\right) \Phi_1^\dagger\Phi_2 -
\left(\delta \mbox{Re}(m_{12}^2) - \ii \delta \mbox{Im}(m_{12}^2)\right)
\Phi_2^\dagger \Phi_1 \notag \\ 
&+ \frac{\delta \lambda_1}{2} \left(\Phi_1^\dagger\Phi_1\right)^ 2 +
  \frac{\delta \lambda_2}{2} \left(\Phi_2^\dagger\Phi_2\right)^2 +
  \delta\lambda_3\left(\Phi_1^\dagger\Phi_1\right)\left(\Phi_2^\dagger\Phi_2\right)
  + \delta \lambda_4\left(\Phi_1^\dagger\Phi_2\right)\left(\Phi_2^\dagger\Phi_1\right)
  \notag \\ 
&+ \frac{1}{2} \left( \delta \mbox{Re}(\lambda_5) + \ii\delta
  \mbox{Im}(\lambda_5) \right) \left(\Phi_1^\dagger\Phi_2\right)^2 +
  \frac{1}{2} \left(\delta \mbox{Re}(\lambda_5) -\ii\delta
  \mbox{Im}(\lambda_5) \right)
  \left(\Phi_2^\dagger\Phi_1\right)^2 \notag \\ 
&+ \delta T_1\left(\zeta_1+\omega_1\right)+\delta
  T_2\left(\zeta_2+\omega_2\right) + \delta
  T_{\text{CP}} \left(\psi_2+\omega_{\text{CP}}\right) + \delta
  T_{\text{CB}}\left(\rho_2+\omega_{\text{CB}}\right) \,.
\end{align}
Using 
\beq
\phi_i \equiv \{ \rho_1, \eta_1, \rho_2, \eta_2, \zeta_1, \psi_1,
\zeta_2, \psi_2 \} \label{eq:vecc2hdm}
\eeq
the 'on-shell' renormalization conditions yield
\beq
\left.\partial_{\phi_i} V^{\text{CT}}\right|_{\phi =\langle \phi^c\rangle_{T=0}} &=&
-\left.\partial_{\phi_i} V^{\text{CW}}\right|_{\phi =\langle \phi^c \rangle_{T=0}} \\
\left.\partial_{\phi_i}\partial_{\phi_j} V^{\text{CT}}
\right|_{\phi=\langle \phi^c \rangle_{T=0}} &=& -\left.\partial_{\phi_i}
\partial_{\phi_j} V^{\text{CW}} \right|_{\phi=\langle\phi^c \rangle_{T=0}}  \,, 
\eeq
with
\beq
\langle \phi^c \rangle_{T=0} = (0,0,0,0,v_1,0,v_2,0) \;,
\eeq
and lead to the counterterms
\begin{align}
\delta m_{11}^2 &= \frac{1}{2} \left[ H^{\CW}_{\zeta_1,\zeta_1} -2H^{\CW}_{\psi_1,\psi_1} - 
H^{\CW}_{\eta_1,\eta_2}\frac{v_2}{v_1} +H^{\CW}_{\zeta_1,\zeta_2}\frac{v_2}{v_1} - 
H^{\CW}_{\rho_1,\rho_1} \right] +v_2^2 t \label{eq:delbeg} \\
\delta m_{22}^2 &= \left[   - \frac{1}{2} \frac{v_1}{v_2} \left( H^{\CW}_{\eta_1,\eta_2} - 
H^{\CW}_{\zeta_1,\zeta_2} \right) + \frac{v_1^2}{v_2^2} \left(H^{\CW}_{\rho_1,\rho_1} -
H^{\CW}_{\psi_1,\psi_1}\right) - \frac{3}{2}H^{\CW}_{\eta_2,\eta_2} + \frac{1}{2}
H^{\CW}_{\zeta_2,\zeta_2} \right] + v_1^2 t \\
\delta\Re \,(m_{12}^2) &= \left[ H^{\CW}_{\eta_1,\eta_2} -
\frac{v_1}{v_2} H^{\CW}_{\psi_1,\psi_1} +  
\frac{v_1}{v_2} H^{\CW}_{\rho_1,\rho_1}\right] + v_1v_2 t \\
\delta \lambda_1 &= \frac{1}{v_1^2} \left[ H^{\CW}_{\psi_1,\psi_1} -H^{\CW}_{\zeta_1,\zeta_1} 
\right]  -\frac{v_2^2}{v_1^2} t\\
\delta \lambda_2 &= \frac{1}{v_2^2} \left[ H^{\CW}_{\eta_2,\eta_2} - H^{\CW}_{\zeta_2,\zeta_2} 
+ \frac{v_1^2}{v_2^2} \left( H^{\CW}_{\psi_1,\psi_1} - H^{\CW}_{\rho_1,\rho_1}\right)\right] -
\frac{v_1^2}{v_2^2} t \\
\delta\lambda_3 &= \frac{1}{v_1v_2} \left[ H^{\CW}_{\eta_1,\eta_2} -H^{\CW}_{\zeta_1,\zeta_2} 
+ \frac{v_1}{v_2} \left(H^{\CW}_{\psi_1,\psi_1} - H^{\CW}_{\rho_1,\rho_1}\right)\right] -t\\
\delta\lambda_4 &= \frac{2}{v_2^2} \left[ H^{\CW}_{\rho_1,\rho_1} - H^{\CW}_{\psi_1,\psi_1} 
\right] +t  \\
\delta\Re \,(\lambda_5) &= t \label{eq:delend} \\
\delta\Im \,(\lambda_5) &= -\frac{2}{v_2^2}H^{\CW}_{\zeta_1,\psi_1} \\
\delta\Im \,(m_{12}^2) &= -\left[ H^{\CW}_{\zeta_1,\psi_2} + 2\frac{v_1}{v_2} 
H^{\CW}_{\zeta_1,\psi_1} \right] \\
\delta T_1 &= H^{\CW}_{\eta_1,\eta_2} v_2 + H^{\CW}_{\rho_1,\rho_1} v_1- N^{\CW}_{\zeta_1} \\
\delta T_2 &= H^{\CW}_{\eta_1,\eta_2} v_1 + H^{\CW}_{\eta_2,\eta_2} v_2 - N^{\CW}_{\zeta_2} \\
\delta T_{\text{CP}} &= \frac{v_1^2}{v_2} H^{\CW}_{\zeta_1,\psi_1} + H^{\CW}_{\zeta_1,\psi_2} v_1 
- N^{\CW}_{\psi_2} \\
\delta T_{\text{CB}} &= -N^{\CW}_{\rho_2} \;,
\end{align}
where we used the abbreviations Eqs.~(\ref{eq:nabbr}) and (\ref{eq:habbr}).
Again, the system of equations is overconstrained. Its one-dimensional
solution space is parametrized by $t \in \mathbb{R}$ which we
have chosen such that 
\begin{align}
\delta \lambda_4 &= 0 \,.
\end{align}
With this choice Eqs.~(\ref{eq:delbeg})--(\ref{eq:delend}) simplify to 
\begin{align}
\delta m_{11}^2 &= \frac{1}{2} \left[ H_{\zeta_1,\zeta_1}^{\CW} +2H_{\psi_1,\psi_1}^{\CW} 
- \frac{v_2}{v_1} \left(H^{\CW}_{\eta_1,\eta_2}-H^{\CW}_{\zeta_1,\zeta_2}\right)
- 5H^{\CW}_{\rho_1,\rho1} \right]\\
\delta m_{22}^2 &= \frac{1}{2} \left[ \frac{v_1}{v_2} \left(H^{\CW}_{\zeta_1,\zeta_2} -
H^{\CW}_{\eta_1,\eta_2}\right) - \frac{v_1^2}{v_2^2}\left(H^{\CW}_{\rho_1,\rho_1}-
H^{\CW}_{\psi_1,\psi_1}\right)- 3H^{\CW}_{\eta_2,\eta_2} + H^{\CW}_{\zeta_2,\zeta_2}\right] \\
\delta\Re \,(m_{12}^2) &= \frac{v_1}{v_2} \left(H^{\CW}_{\psi_1,\psi_1}  -H^{\CW}_{\rho_1,\rho_1}
\right) +H^{\CW}_{\eta_1,\eta_2} \\
\delta \lambda_1 &= \frac{1}{v_1^2} \left( 2H^{\CW}_{\rho_1,\rho_1} -H^{\CW}_{\psi_1,\psi_1} -
H^{\CW}_{\zeta_1,\zeta_1}\right) \\
\delta \lambda_2 &= \frac{1}{v_2^2} \left[ \frac{v_1^2}{v_2^2}\left(H^{\CW}_{\rho_1,\rho_1}
-H^{\CW}_{\psi_1,\psi_1}\right)+ H^{\CW}_{\eta_2,\eta_2} -H^{\CW}_{\zeta_2,\zeta_2}\right] \\
\delta \lambda_3 &= \frac{1}{v_1v_2^2} \left[ \left(H^{\CW}_{\rho_1,\rho_1} -
H^{\CW}_{\psi_1,\psi_1}\right) v_1 + \left(H^{\CW}_{\eta_1,\eta_2} -H^{\CW}_{\zeta_1,\zeta_2}
\right)v_2 \right] 
\\
\delta \lambda_4 &= 0 \\
\delta \Re \,(\lambda_5) &= \frac{2}{v_2^2} \left(H^{\CW}_{\psi_1,\psi_1} -
H^{\CW}_{\rho_1,\rho_1}\right)  \;,
\end{align}
where we have applied the identities needed for the consistent solution,
\begin{align}
0 &= H^{\CW}_{\eta_1,\eta_1} - H^{\CW}_{\rho_1,\rho_1} \\
0 &= H^{\CW}_{\eta_1,\eta_2} - H^{\CW}_{\rho_1,\rho_2} \\
0 &= H^{\CW}_{\eta_2,\eta_2} - H^{\CW}_{\rho_2,\rho_2} \\
0 &= H^{\CW}_{\psi_1,\psi_2} - H^{\CW}_{\eta_1,\eta_2} +\frac{v_1}{v_2} \left( 
H^{\CW}_{\psi_1,\psi_1} -H^{\CW}_{\rho_1,\rho_1} \right) \\
0 &= H^{\CW}_{\psi_2,\psi_2} - H^{\CW}_{\eta_2,\eta_2} +\frac{v_1^2}{v_2^2} \left( 
H^{\CW}_{\rho_1,\rho_1} - H^{\CW}_{\psi_1,\psi_1}\right) \\
0 &= H^{\CW}_{\zeta_1,\psi_2} v_2 + v_1H^{\CW}_{\zeta_1,\psi_1} v_1 + N^{\CW}_{\psi_1} \\
0 &= H^{\CW}_{\rho_1,\eta_2} - H^{\CW}_{\zeta_1,\psi_2} -\frac{v_1}{v_2} H^{\CW}_{\zeta_1,\psi_1} \\
0 &= H^{\CW}_{\zeta_1,\psi_2} + H^{\CW}_{\eta_1,\rho_2} + \frac{v_1}{v_2} 
H^{\CW}_{\zeta_1,\psi_1} \\
0 &= H^{\CW}_{\psi_1,\zeta_2} + H^{\CW}_{\zeta_1,\psi_2} \\
0 &= \frac{v_1^2}{v_2^2} H^{\CW}_{\zeta_1,\psi_1} + H^{\CW}_{\zeta_2,\psi_2} \;.
\end{align}
Note that $\delta T_{\text{CB}}$ related to the charge breaking VEV turns out to
be zero as we do not have a charge-breaking vacuum. \s

The C2HDM is parametrized by nine independent parameters. In a
physics-inspired basis the masses are part of the input, in the 'parametric'
basis, used in the code, the input parameters in addition to the SM
VEV hard-coded in the program, are, in the order required by the
program,  
\beq
\mbox{type} \;, \; \lambda_1 \;, \;\lambda_2 \;, \;\lambda_3 \;, \;\lambda_4 \;, \;
\mbox{Re}\lambda_5 \;, \;\mbox{Im} (\lambda_5) \;, \;\mbox{Re}
(m_{12}^2) \;, \; \tan\beta \;.
\eeq
%

\noindent		
By setting $\mbox{type}=1,2,3$ or 4, the user chooses the C2HDM type. 
The parameters $m_{11}^2,m_{22}^2$ and $\mbox{Im}(m_{12}^2)$ are
obtained from the minimum conditions  
\begin{align}
m_{11}^2 &= \mbox{Re} (m_{12}^2) \frac{v_2}{v_1} -\frac{v_1^2}{2}\lambda_1^2 
- \frac{v_2^2}{2} \left(\lambda_3+\lambda_4+\mbox{Re} (\lambda_5) \right) \\ 
m_{22}^2 &= \mbox{Re} (m_{12}^2) \frac{v_1}{v_2} - \frac{v_2^2}{2}\lambda_2^2 
- \frac{v_1^2}{2} \left(\lambda_3+\lambda_4+\mbox{Re} (\lambda_5) \right) \\
\mbox{Im} (m_{12}^2) &= \frac{v_1v_2}{2} \mbox{Im} (\lambda_5) \,.
\end{align}

\subsection{The N2HDM}
The N2HDM is built from the CP-conserving 2HDM with a softly broken
$\mathbb{Z}_2$ symmetry upon extension by a singlet field
$\Phi_S$. If the latter does not acquire a VEV, we have a dark matter candidate
\cite{He:2008qm}. Here, we let the singlet field have a non-vanishing VEV. (For
the phenomenology of the N2HDM with a singlet VEV, see \cite{Chen:2013jvg} with
and \cite{phenocomp, Muhlleitner:2016mzt} without any
approximations. The NLO electroweak corrected 
N2HDM and in particular its renormalization has been presented in
\cite{Krause:2017mal}.) The tree-level potential of the N2HDM is given
by
\beq
V^{(0)} &=& m_{11}^2 \Phi_1^\dagger \Phi_1 + m_{22}^2\Phi_2^\dagger \Phi_2 -
m_{12}^2\left(\Phi_1^\dagger\Phi_2 +\Phi_2^\dagger \Phi_1\right) +
\frac{\lambda_1}{2}\left(\Phi_1^\dagger\Phi_1\right)^ 2 +
\frac{\lambda_2}{2}\left(\Phi_2^\dagger\Phi_2\right)^2 \nonumber \\
&&+\lambda_3\left(\Phi_1^\dagger\Phi_1\right)\left(\Phi_2^\dagger\Phi_2\right)
+\lambda_4\left(\Phi_1^\dagger\Phi_2\right)\left(\Phi_2^\dagger\Phi_1\right)
+ \frac{\lambda_5}{2} \left[\left(\Phi_1^\dagger\Phi_2\right)^2 +
\left(\Phi_2^\dagger\Phi_1\right)^2\right] \nonumber \\
&&+ \frac{1}{2} m_S^2 \Phi_S^2 + \frac{\lambda_6}{8} \Phi_S^4 +
\frac{\lambda_7}{2} \left(\Phi_1^\dagger\Phi_1\right)\Phi_S^2 +
\frac{\lambda_8}{2} \left(\Phi_2^\dagger\Phi_2\right)\Phi_S^2 \;,
\eeq
where the first two lines describe the 2HDM part of the N2HDM and the
last line is the contribution of the singlet field $\Phi_S$. The
potential obeys two $\mathbb{Z}_2$ symmetries. The first one, named
$\mathbb{Z}_2$, is the trivial generalization of the usual 2HDM
$\mathbb{Z}_2$ symmetry to the N2HDM,
\beq
\Phi_1 \to \Phi_1 \;, \quad \Phi_2 \to - \Phi_2 \;, \quad \Phi_S \to 
\Phi_S \;, 
\eeq
and is softly broken by the term proportional to $m_{12}^2$. Its
extension to the Yukawa sector ensures the absence of FCNC and implies
different types of N2HDM that are the same as in the 2HDM, summarized
in Table \ref{tab:chargassign}. The second one, named
$\mathbb{Z}_2^\prime$, is given by 
\beq
\Phi_1 \to \Phi_1 \;, \quad \Phi_2 \to \Phi_2 \;, \quad \Phi_S \to 
-\Phi_S \;, 
\eeq
and is not explicitly broken. For a non-vanishing VEV of $\Phi_S$ as
allowed here, there is mixing among all CP-even neutral scalars. This
is also the case if $m_{12}^2=0$, which will not be considered here,
however. After EWSB, the doublets and the singlet field acquire VEVs
about which they can be expanded as (allowing for the most general
vacuum configuration with CP- and (unphysical) CB-violating VEVs),
\beq
\Phi_1 &=& \frac{1}{\sqrt{2}} \begin{pmatrix}
\rho_1 + \ii \eta_1 \\ \zeta_1 + \bar{\omega}_1 + \ii \psi_1
	\end{pmatrix} \\
\Phi_2 &=& \frac{1}{\sqrt{2}} \begin{pmatrix}
\rho_2 + \bar{\omega}_{\text{CB}} + \ii \eta_2 \\ \zeta_2 + \bar{\omega}_2 + \ii \left(\psi_2 + \bar{\omega}_{\text{CP}} \right)
\end{pmatrix} \\
\Phi_S &=& \bar{\omega}_S + \rho_S \;.
\eeq
The diagonalization of the mass matrix of the neutral scalar fields,
obtained after EWSB from the second derivative of the potential with respect to
these fields, leads to three neutral physical Higgs states, $H_{1}$,
$H_2$ and $H_3$ that are ordered by ascending mass, {\it
  i.e.}~$m_{H_1} \le m_{H_2} \le m_{H_3}$. The CP-odd and the charged
sector do not change with respect to the real 2HDM, and we have a
pseudoscalar Higgs $A$ and two charged Higgs states $H^\pm$. The N2HDM
counterterm potential reads
\beq
V^{\text{CT}} &=& \delta m_{11}^2 \Phi_1^\dagger \Phi_1 + \delta
m_{22}^2\Phi_2^\dagger \Phi_2 - \delta
m_{12}^2\left(\Phi_1^\dagger\Phi_2 +\Phi_2^\dagger \Phi_1\right) + 
\frac{\delta\lambda_1}{2}\left(\Phi_1^\dagger\Phi_1\right)^ 2 +
\frac{\delta\lambda_2}{2}\left(\Phi_2^\dagger\Phi_2\right)^2 \nonumber \\
&&+\delta\lambda_3\left(\Phi_1^\dagger\Phi_1\right)\left(\Phi_2^\dagger\Phi_2\right)
+\delta\lambda_4\left(\Phi_1^\dagger\Phi_2\right)\left(\Phi_2^\dagger\Phi_1\right)
+ \frac{\delta\lambda_5}{2} \left[\left(\Phi_1^\dagger\Phi_2\right)^2 +
\left(\Phi_2^\dagger\Phi_1\right)^2\right] \nonumber \\
&&+ \frac{1}{2} \delta m_S^2 \Phi_S^2 + \frac{\delta \lambda_6}{8} \Phi_S^4 +
\frac{\delta \lambda_7}{2} \left(\Phi_1^\dagger\Phi_1\right)\Phi_S^2 +
\frac{\delta \lambda_8}{2} \left(\Phi_2^\dagger\Phi_2\right)\Phi_S^2 \nonumber \\
&&+ \delta T_1 (\zeta_1 + \omega_1) + \delta T_2 (\zeta_2 + \omega_2) + \delta T_{\text{CP}} (\psi_2 + \omega_{\text{CP}}) \nonumber \\
&& + \delta T_{\text{CB}} (\rho_2 + \omega_{\text{CB}}) + \delta T_S ( \rho_S + \omega_S ) \;.
\eeq
Using 
\beq
\phi_i \equiv \{ \rho_1, \eta_1, \rho_2, \eta_2, \zeta_1, \psi_1,
\zeta_2, \psi_2, \rho_S \}
\eeq
the 'on-shell' renormalization conditions yield
\beq
\left.\partial_{\phi_i} V^{\text{CT}}\right|_{\phi =\langle \phi^c\rangle_{T=0}} &=&
-\left.\partial_{\phi_i} V^{\text{CW}}\right|_{\phi =\langle \phi^c \rangle_{T=0}} \\
\left.\partial_{\phi_i}\partial_{\phi_j} V^{\text{CT}}
\right|_{\phi=\langle \phi^c \rangle_{T=0}} &=& -\left.\partial_{\phi_i}
\partial_{\phi_j} V^{\text{CW}} \right|_{\phi=\langle\phi^c \rangle_{T=0}}  \,, 
\eeq
with
\beq
\langle \phi^c \rangle_{T=0} = (0,0,0,0,v_1,0,v_2,0,v_S) \;,
\eeq
and lead to the counterterms
\begin{align}
\delta m_{11}^2 =& \frac{1}{2} \left[ \frac{v_s}{v_1} H^{\CW}_{\rho_1,\rho_S} + \frac{v_2}{v_1} 
\left(H^{\CW}_{\rho_1,\rho_2} - H^{\CW}_{\psi_1,\psi_2} \right) + 2H^{\CW}_{\psi_1,\psi_1} -
5H^{\CW}_{\psi_1,\psi_1} + H^{\CW}_{\rho_1,\rho_1}  \right] + t_H v_2^2 \\
\delta m_{22}^2 =& \frac{1}{2} \left[ \frac{v_s}{v_2} H^{\CW}_{\rho_2,\rho_S} + 
H^{\CW}_{\rho_2,\rho_2} - 3H^{\CW}_{\psi_2,\psi_2}  + \frac{v_1}{v_2} \left(
H^{\CW}_{\rho_1,\rho_2}  - H^{\CW}_{\psi_1,\psi_2} \right) + 5\frac{v_1^2}{v_2^2}
\left(H^{\CW}_{\psi_1,\psi_1}-H^{\CW}_{\psi_1,\psi_1}\right) \right] + t_H v_1^2 \\
\delta m_{12}^2 &= H^{\CW}_{\psi_1,\psi_2} + \frac{v_1}{v_2}\left(H^{\CW}_{\psi_1,\psi_1} -
H^{\CW}_{\psi_1,\psi_1} \right) + t_H v_1v_2 \\
\delta \lambda_1 &= \frac{1}{v_1^2} \left( 2H^{\CW}_{\psi_1,\psi_1} - H^{\CW}_{\psi_1,\psi_1} -
H^{\CW}_{\rho_1,\rho_1}\right) - t_H \frac{v_2^2}{v_1^2} \\
\delta \lambda_2 &= \frac{1}{v_2^2}\left(H^{\CW}_{\psi_2,\psi_2}-H^{\CW}_{\rho_2,\rho_2}\right)
+ 2\frac{v_1^2}{v_2^4} \left(H^{\CW}_{\psi_1,\psi_1}-H^{\CW}_{\psi_1,\psi_1}\right)
- t_H \frac{v_1^2}{v_2^2} \\
\delta \lambda_3 &= \frac{1}{v_2^2} \left( H^{\CW}_{\psi_1,\psi_1} - H^{\CW}_{\psi_1,\psi_1} \right) 
+ \frac{1}{v_1v_2} \left( H^{\CW}_{\psi_1,\psi_2} - H^{\CW}_{\rho_1,\rho_2} \right) -t_H \\
\delta \lambda_4 &= t_H \\
\delta \lambda_5 &= \frac{2}{v_2^2} \left(H^{\CW}_{\psi_1,\psi_1} - 2H^{\CW}_{\psi_1,\psi_1}
\right) + t_H \\
\delta m_S^2 &= \frac{1}{2} \left( H^{\CW}_{\rho_S,\rho_S} + \frac{v_2}{v_s} H^{\CW}_{\rho_2,\rho_S} +
\frac{v_1}{v_s} H^{\CW}_{\rho_1,\rho_S} - \frac{3}{v_S} N^{\CW}_{\rho_S} \right) - t_S \frac{3}{2v_s} \\
\delta \lambda_6 &= \frac{1}{v_s^3} \left(N^{\CW}_{\rho_S} - v_s H^{\CW}_{\rho_S,\rho_S} \right) 
- t_S \frac{1}{v_s^3}\\
\delta \lambda_7 &=  - \frac{1}{v_sv_1} H^{\CW}_{\rho_1,\rho_S} \\
\delta \lambda_8 &= - \frac{1}{v_sv_2} H^{\CW}_{\rho_2,\rho_S} \\
\delta T_1 &= H^{\CW}_{\psi_1,\psi_1}v_1 + H^{\CW}_{\psi_1,\psi_2} v_2 - N^{\CW}_{\rho_1} \\
\delta T_2 &= \frac{v_1^2}{v_2} \left(H^{\CW}_{\psi_1,\psi_1} - H^{\CW}_{\psi_1,\psi_1}\right) 
+ H^{\CW}_{\psi_1,\psi_2} v_1 + H_{\psi_2,\psi_2} v_2 - N^{\CW}_{\zeta_2} \\
\delta T_S &= t_S \\
\delta T_{\text{CP}} &= \frac{v_1^2}{v_2} H^{\CW}_{\rho_1,\psi_1} + H^{\CW}_{\rho_1,\psi_2} v_1 
- N^{\CW}_{\psi_2} \\
\delta T_{\text{CB}} &= -N^{\CW}_{\zeta_2} \;,
\end{align}
where we used the abbreviations Eqs.~(\ref{eq:nabbr}) and (\ref{eq:habbr}).
The overconstrained system of equations leads to a two-dimensional
solution space parametrized by $t_H, t_S \in \mathbb{R}$ that we set
in the code to 
\beq
t_H = t_S = 0 \;.
\eeq
The identities to be applied to solve the system of equations are the
same as in the R2HDM and given by
\begin{align}
0 &= H^{\CW}_{\eta_1,\eta_1} - H^{\CW}_{\rho_1,\rho_1} \\
0 &= H^{\CW}_{\eta_1,\eta_2} - H^{\CW}_{\rho_1,\rho_2} \\
0 &=H^{\CW}_{\eta_2,\eta_2} - H^{\CW}_{\rho_2,\rho_2} \\
0 &= H^{\CW}_{\psi_1,\psi_2} - H^{\CW}_{\eta_1,\eta_2} + \frac{v_1}{v_2}
    \left( H^{\CW}_{\psi_1,\psi_1} - H^{\CW}_{\rho_1,\rho_1} \right) 
\\
0 &= H^{\CW}_{\psi_2,\psi_2} - H^{\CW}_{\eta_2,\eta_2} +
    \frac{v_1^2}{v_2^2} \left( H^{\CW}_{\rho_1,\rho_1} -
    H^{\CW}_{\psi_1,\psi_1}\right) \,.
\end{align}
As a charge-breaking vacuum is unphysical, $\delta
  T_{\text{CB}}$ always turns out to be zero as it should.
The program code requires (in addition to the SM VEV that is
hard-coded) the 'parametric' input parameters for the 
N2HDM to be given in the order 
\beq
\mbox{type} \; , \; \lambda_1 \; , \; \lambda_2 \; , \; \lambda_3 \; , \; \lambda_4 \; ,
\; \lambda_5 \; , \; \lambda_6 \; , \; 
\lambda_7 \; , \; \lambda_8 \; , \;   v_S \; , \; \tan\beta \; ,
\; m_{12}^2 \;.
\eeq
In the first entry, the user has to specify the N2HDM type. 
Note that the minimum conditions lead to the
following relations among the parameters
\beq
\frac{v_2}{v_1} m_{12}^2 - m_{11}^2 &=& \frac{1}{2} (v_1^2 \lambda_1 +
v_2^2 \lambda_{345} + v_S^2 \lambda_7) \label{eq:n2hdmmin1} \\
\frac{v_1}{v_2} m_{12}^2 - m_{22}^2 &=& \frac{1}{2} (v_1^2 \lambda_{345} +
v_2^2 \lambda_2 + v_S^2 \lambda_8) \label{eq:n2hdmmin2} \\
- m_S^2 &=& \frac{1}{2} (v_1^2 \lambda_7 + v_2^2 \lambda_8 + v_S^2
\lambda_6) \;, \label{eq:n2hdmmin3}
\eeq
with
\beq
\lambda_{345} \equiv \lambda_3 + \lambda_4 + \lambda_5 \;.
\label{eq:l345}
\eeq

\section{Installation \label{sec:install}}
\paragraph*{Download}
The program can be downloaded from \download\ . After extracting the
zip archive in the directory chosen by the user, to which we will from now
on refer as {\tt \$BSMPT}, there will be several subfolders. These are: 
\vspace*{0.2cm}

\begin{tabular}{ll}
{\tt docs} & The {\tt docs} folder contains the documentation as html.\\[0.1cm]
{\tt example} & Here we put sample input files as well as the corresponding results
          produced \\ & by the different executables (see
                        below). \\[0.1cm]
{\tt manual} & This subfolder contains a copy of this paper which is
               kept up to date \\ & with changes in the code. 
         Additionally, we include the {\tt changelog} \\ & file
         documenting corrected bugs and modifications of the
         program. \\[0.1cm]
{\tt sh} & Here we put the script to install the libraries and to create the
     makefile. Here \\ & we also provide the python files
                    prepareData\_XXX.py (XXX= R2HDM,  \\ 
    &   C2HDM, N2HDM) that can be used to order the 
         data sample accordingly to \\
    &   the input requirements. \\[0.1cm]
{\tt src} & This subfolder contains the source files of the code and 
            is structured in three 
\\ & subfolders. 
\end{tabular}

\noindent
The subfolders of {\tt src} contain the following files:
\vspace*{0.2cm}

\begin{tabular}{ll}
\hspace*{-0.32cm} {\tt src/minimizer} & Here the source files for the
                       minimization routines are stored. \\
\hspace*{-0.32cm} {\tt src/models} & This directory contains the implemented
                    models. \\ &  If a new model is added it must 
                    be placed in this folder. \\ & There is also a template
                    class with instructions on how \\ & to add a new
                    model. Furthermore, there is the file \\ & {\tt SMparam.h}
                    with the Standard Model parameters. \\
\hspace*{-0.32cm} {\tt src/prog} & This directory contains the source code for
                  the executables.
\end{tabular}

\paragraph*{Required libraries}
For {\tt BSMPT} to work the following three libraries are needed:
\begin{itemize}
\item[$\ast$] The GNU Scientific Library ({\tt GSL}) \cite{GSL_Manual} is
  assumed to be installed in PATH. {\tt GSL} is
  required for the calculation of the Riemann-$\zeta$ functions, the 
  double factorial and for the minimization.  
\item[$\ast$] The {\tt Eigen3} library \cite{eigenweb} is downloaded
  during the installation process of {\tt BSMPT}. {\tt Eigen3} is used for all
  the matrix calculations. 
\item[$\ast$] The {\tt libcmaes} library \cite{CMAES} is required
  for the minimization and is installed during the installation process.
\end{itemize}

\paragraph*{Compilation}
The compilation requires a {\tt C++} and {C} compiler
  that support the {\tt C++11} standard. For the {\tt C++} compiler we
  recommend {\tt g++-7\footnote{Although
  earlier compiler versions can also be used, we strongly recommend to
use {\tt g++-7} as it significantly reduces the computation time.} and
for the {\tt C} compiler we recommend {\tt gcc-7}.}
After that, the following steps have to be performed:
\begin{enumerate}
\item Go to the folder {\tt \home/sh} and call \\[0.1cm]
\begin{kasten*}{}
\vspace*{-0.2cm}
./InstallLibraries.sh \dash\dash lib=\PathToYourLib{}  \dash\dash CXX=C++Compiler  \dash\dash CC=CCompiler
\end{kasten*} \\
where '\PathToYourLib', is the absolute path in which
{\tt Eigen3} and {\tt libcmaes} will be installed by this script. 
\item To generate the Makefile in the {\tt \home}\, folder call \\[0.1cm]
\begin{kasten*}{}
\vspace*{-0.2cm}
		./autogen.sh \dash\dash lib=\PathToYourLib{}  \dash\dash CXX=C++Compiler
\end{kasten*} 
\item Go back to {\tt \home}\, and call \\[0.1cm]
\begin{kasten*}{}
\vspace*{-0.2cm}
		make
\end{kasten*} \\
which will generate the executables {\tt BSMPT}, {\tt
  CalcCT}, {\tt NLOVEV}, {\tt TripleHiggsNLO} and {\tt VEVEVO}.
\end{enumerate}
After that go to the folder {\tt \PathToYourLib/libcmaes} and check if there
is either the folder {\tt lib} or {\tt lib64}. Then \\[0.1cm]  
\begin{kasten*}{}
\vspace*{-0.2cm}
export LD\_LIBRARY\_PATH=\$LD\_LIBRARY\_PATH:\PathToYourLib/libcmaes/LIB
\end{kasten*} \\
has to be executed where 'LIB' is either {\tt lib64} or {\tt lib}, depending
        on which folder exists in {\tt \PathToYourLib/libcmaes}. This can
        also be added to {\tt bashrc} so that it is loaded
        automatically with every new terminal that is opened.

\section{Executables \label{sec:executables}}
In this section we will briefly describe the executables that are
generated by the makefile. We begin with the definition of the input
parameters that are used by all executables:
\begin{itemize}
\item {\it Model} is the parameter by which the model is selected. The
  CP-violating 2HDM (0), the
  CP-conserving 2HDM (1) and the 
  CP-conserving N2HDM (2), as introduced in
  Section~\ref{sec:implmodels}, are already implemented.  
\item {\it Inputfile} sets the path and the name of the input file.
  In the input file, the programs expect the first line
  to be a header with the column names. Every
  following line then corresponds to the input of one particular
  parameter point. The parameters are required to be those of
  the Lagrangian in the interaction basis. If a different format 
  for the input parameters is desired one needs to adapt the function
  {\tt ReadAndSet} in the corresponding model file in
  {\tt \home/src/models}. For the format of the input files of the
  already implemented models, we refer to the corresponding
  subsections in Sec.~\ref{sec:implmodels}. Note, that the program
  expects the input parameters to be separated by a tabulator. In the
  folder {\tt \$BSMPT/sh/}, we provide {\tt python} scripts that prepare
  the data accordingly.
\item {\it Outputfile} sets the path and the name of
  the generated output file. We note, that the 
  program does not create new folders so that it has to be made sure that the
  folder for the output file already exists. 
\end{itemize}
If in the thermal corrections the Parwani method
Eqs.~(\ref{eq:Parwani1}), (\ref{eq:Parwani2}), should be used instead of the Arnold
Espinosa method, Eqs.~(\ref{eq:AE1}), (\ref{eq:AE2}), the variable 
'C\_UseParwani' in line 132 of the file 
{\tt \home/src/models/ClassPotentialOrigin.h} has to be changed to
'true'. Afterwards, in {\tt \home}\, the commands
'make clean' and subsequently 'make' have to be executed. 

\subsection{BSMPT}
{\tt BSMPT} is the executable of the main program. It
calculates the EWPT for the parameter point(s) given in the input
file. It is executed through the command line  \\[0.1cm]
\begin{kasten*}{BSMPT}
./bin/BSMPT Model Inputfile Outputfile LineStart LineEnd 
\end{kasten*} \\
The user has to specify the model, the name and path of the input file
and the name and path of the output file through {\it Model}, {\it Inputfile} and {\it
  Outputfile}, respectively. By {\it LineStart} and {\it LineEnd} the
numbers of the lines in the 
input file are specified where the set of parameter points starts and
ends for which the program performs the calculations. Each line
corresponds to one parameter point. Note that the
  first line of your data (the line with the legend) has the number
  1. The code reads in a line from the input file, calculates the EWPT for
this parameter point and then writes out the line in the output file,
{\it i.e.} the information on the parameter point, and appends the
results of the calculations. These are $v_c$, $T_c$, $v_c/T_c$ and
the individual VEVs at $T_c$, {\it i.e.}, $\bar{\omega}_k (T_c)$
($k=1,...,n_v$). It also extends the legend from the input file by adding the entries
for the output. \s

Only results for those
points are written out for which $v_c/T_c > 1$. If the
check should not be against 1 but against a different value the
constant 'C\_PT' in line 154 
of the file \\ {\tt \home/src/model/ClassPotentialOrigin.h} has to be
changed to the desired value. Afterwards, in {\tt \home}\, the commands
'make clean' and subsequently 'make' have to be executed. 

\subsection{CalcCT}
{\tt CalcCT} is the executable for the calculation of the 
counterterms for a given parameter point. It is executed through the
command line \\[0.1cm]
\begin{kasten*}{CalcCT}
./bin/CalcCT Model Inputfile Outputfile LineStart LineEnd
\end{kasten*}\\
in which the user first has to specify the model, the name and path of the
input file and the name and path of the output file through {\it
  Model}, {\it Inputfile} and {\it Outputfile},
respectively. Furthermore, the line numbers of  
the start and end parameter point have to be specified. For each line,
{\it i.e.}~each parameter point, the various counterterms of the model
are calculated. They are written out in the output file which contains
a copy of the parameter point and appended to it in the same line the
results for the counterterms. The first line of the output file
contains the legend describing the entries of the various columns.

\subsection{NLOVeV}
{\tt NLOVeV} is the executable calculating the 
global minimum of the loop-corrected effective potential at $T=0$~GeV
for every point between the lines {\it LineStart} and {\it LineEnd} to
be specified in the command line for the execution of the program: \\[0.1cm]
\begin{kasten*}{NLOVeV}
./bin/NLOVeV Model Inputfile Outputfile LineStart LineEnd
\end{kasten*}\\
The model, the name and path of the input file and the name and path
of the output file are set through {\it Model}, {\it Inputfile} and {\it
  Outputfile}, respectively. The output file contains the information
on the parameter point to which the computed values at zero
temperature of the NLO VEVs (in GeV) are appended in the same line, namely $v(T=0)$
and the individual VEVs $\bar{\omega}_k (T=0) \equiv v_k$
($k=1,...,n_v$). The first line of the output file again details the
entries of the various columns. Note, that it can happen that
the global minimum $v(T=0)$, that is obtained from the NLO effective potential, 
is not equal to $v=246.22$~GeV any more. By writing out also
$v(T=0)$ the user can check for this phenomenological constraint.

\subsection{TripleHiggsCouplingsNLO}
{\tt TripleHiggsCouplingsNLO} is the executable of the program that
calculates the triple Higgs couplings, derived from the third
derivative of the potential with respect to the Higgs fields, for every point between
the lines {\it LineStart} and {\it LineEnd} to be specified in the
command line: \\[0.1cm] 
\begin{kasten*}{TripleHiggsCouplingsNLO}
./bin/TripleHiggsNLO Model Inputfile Outputfile LineStart LineEnd
\end{kasten*}\\
The model, the name and path of the input file and the name and path
of the output file are set through {\it Model}, {\it Inputfile} and {\it
  Outputfile}, respectively. The output file contains the trilinear
Higgs self-couplings derived from the tree-level potential, the
counterterm potential and the Coleman-Weinberg potential at $T=0$ for
all possible Higgs field combinations. The total NLO trilinear Higgs self-couplings are
then given by the sum of these three contributions. The first line
  of the output file describes the entries of the various columns.

\subsection{VEVEVO}
{\tt VEVEVO} is the executable of the program that calculates the
temperature evolution of the VEVs for a given parameter point. It is
performed through the command line \\[0.1cm]
\begin{kasten*}{VEVEVO}
./bin/VEVEVO Model Inputfile Outputfile Line Tempstart Tempstep Tempend
\end{kasten*} \\
Again, the model, the name and path of the input file and the name and
path of the output file have to be specified through {\it Model}, {\it Inputfile} and {\it
  Outputfile}, respectively. Furthermore, 
\begin{itemize}
\item {\it Line} is the line number of the parameter point for which
  the evolution shall be calculated.
\item {\it Tempstart} is the starting value of the temperature in GeV.
\item {\it Tempstep} is the step size of the temperature evolution for
  which the VEVs are to be calculated.
\item {\it Tempend} is the end value of the temperature interval, in
  which the potential should be minimized. 
\end{itemize}
The output file contains the data for $T$ and the corresponding values of
$v$ and of the individual VEVs, {\it i.e.}~$\bar{\omega}_k (T)$
($k=1,...,n_v$). The first line of the output file is devoted to the
legend that specifies the entries of the various columns. 
Note, that the program does not check whether the individual VEVs at the various
temperatures are positive or not but just writes out the results of the
numerical minimizer, and therefore the signs of the individual VEVs can
flip. \s

\begin{figure}[t]
\centering
\subfigure[Evolution of $v$.]{\includegraphics[scale=0.5]{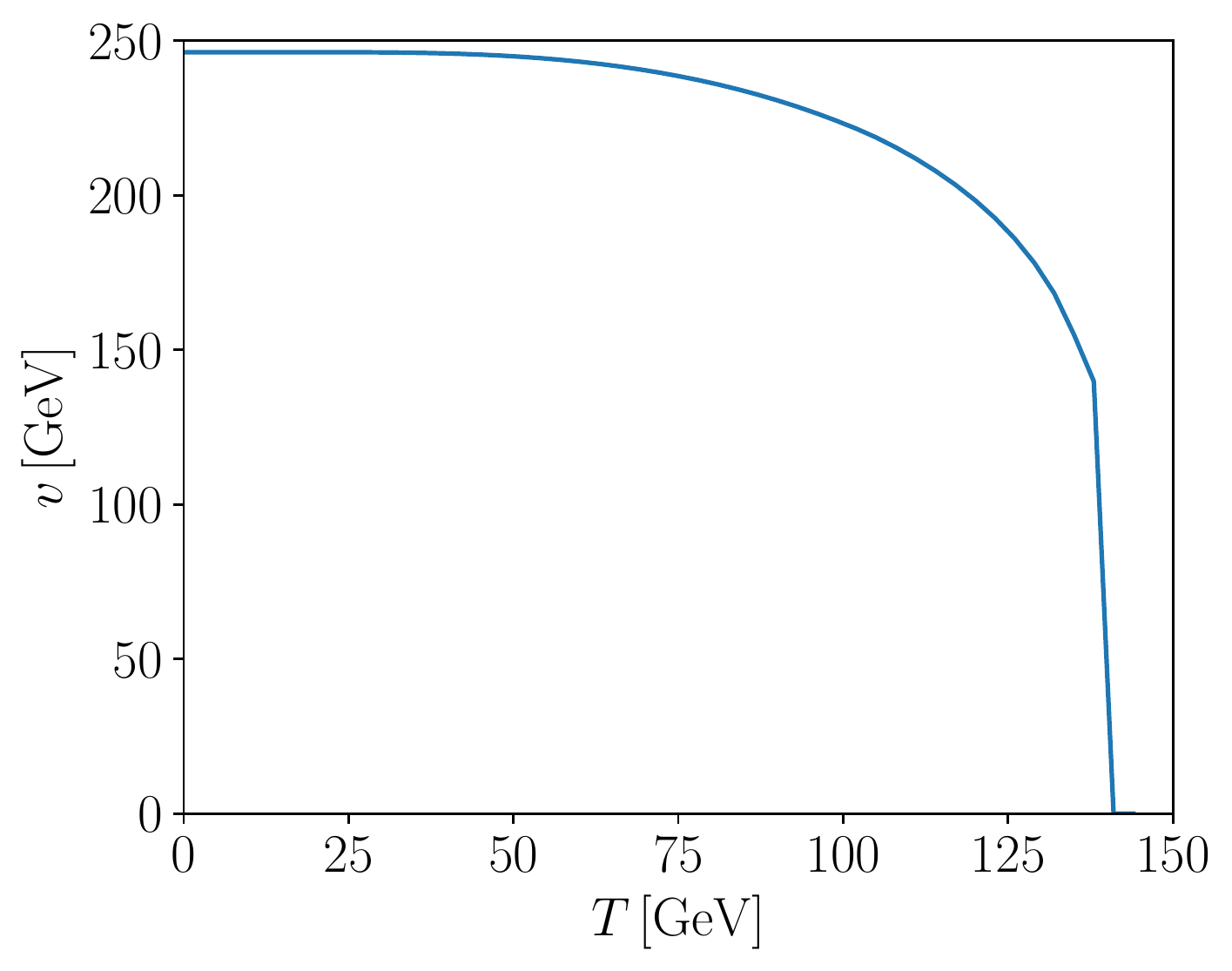}} \qquad
\subfigure[Evolution of the doublet VEVs
$\bar{\omega}_k$
($k=1,2$).]{\includegraphics[scale=0.5]{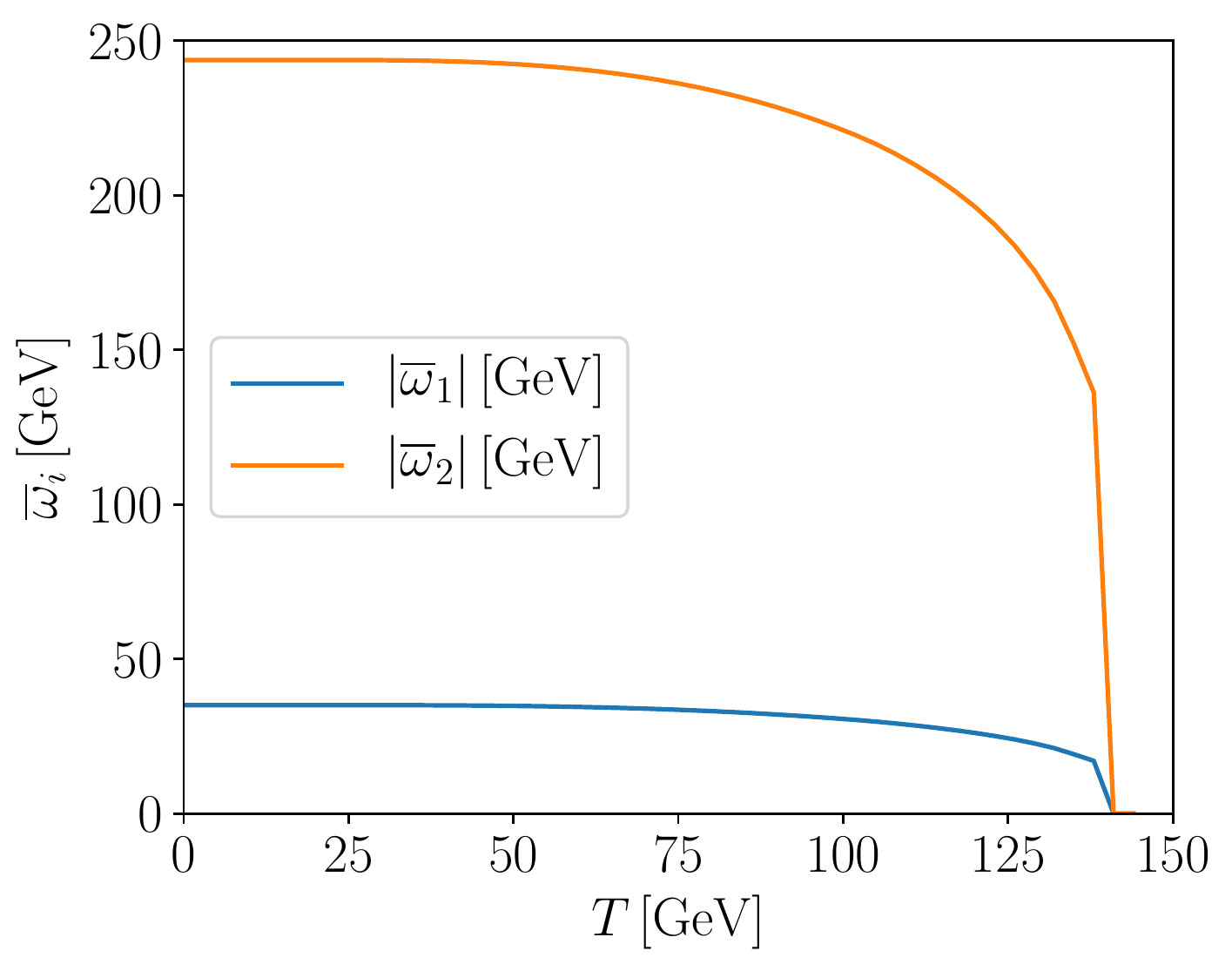}} \\	 
\subfigure[Evolution of the CP-violating phase $\tan(\varphi) =
\bar{\omega}_{\text{CP}}/\bar{\omega}_2$ of the second
doublet.]{\includegraphics[scale=0.5]{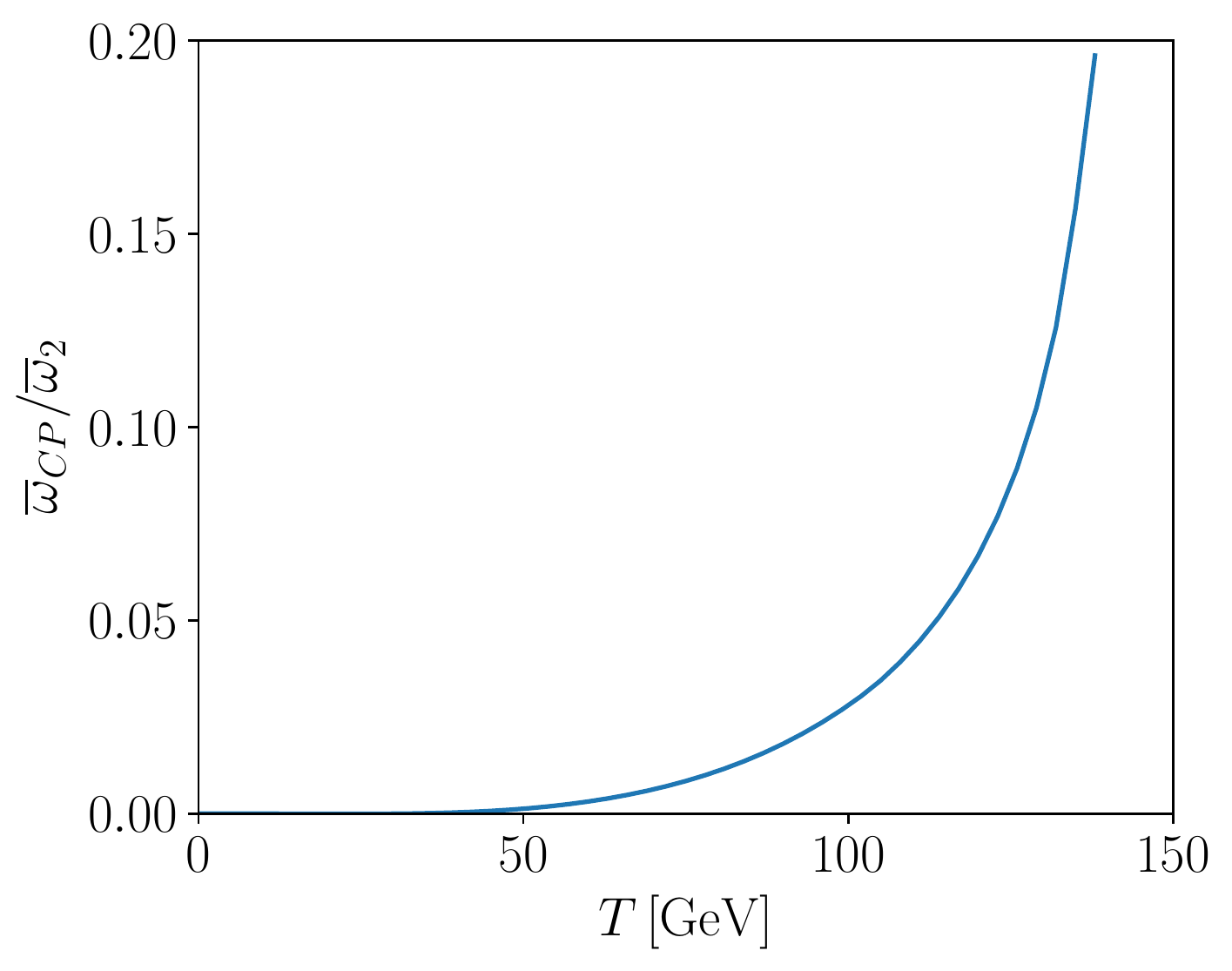}} \qquad 
\subfigure[Evolution of the CP-violating VEV
$\bar{\omega}_{\text{CP}}$ and the charge-breaking VEV
$\bar{\omega}_{\text{CB}}$.]{\includegraphics[scale=0.5]{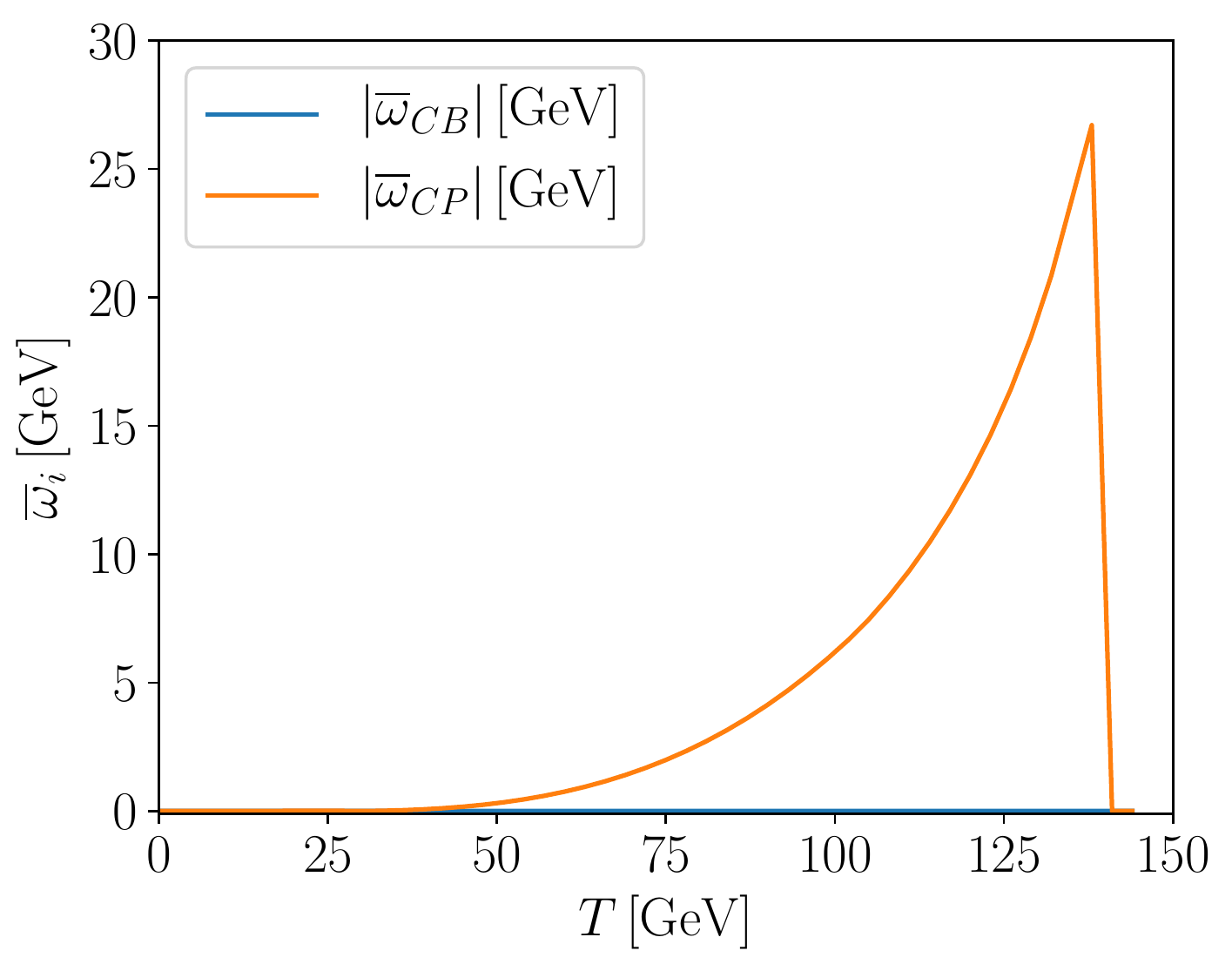}}
\caption{The temperature evolution obtained by {\tt VEVEVO} for the
  C2HDM parameter point Eq.~(\ref{eq:c2hdmex}).}
\label{fig:PlotGen}
\end{figure}

An example for the temperature evolution of a specific parameter point in the C2HDM,
described in section \ref{sec:C2HDM}, is depicted in
Fig.~\ref{fig:PlotGen}. The parameter point is given by the input
values
\beq
\begin{array}{lcllcl}
\mbox{type} & = & 1 & \; \tan\beta & = & 6.94743 \\
 \lambda_1 & = & 1.2248193823 & \; 
\lambda_2 & = & 0.299419454432  \\
\lambda_3 & = & -0.514319430337 & \; \lambda_4 & = & 4.07718269395
\\
\mbox{Re} (\lambda_5) & = & -3.84704455054 & \; \mbox{Im} (\lambda_5) & = & -1.0875150879
  \\
\mbox{Re} (m_{12}^2) & = & 8044.09 \,\mathrm{GeV}^2 \;.
\end{array}  \label{eq:c2hdmex}
\eeq
This implies the Higgs boson masses 
\beq
\begin{array}{lcllcl}
m_{H_1} &=& 125.09 \,\mathrm{GeV} & \; 
m_{H_2} &=& 236.989 \,\mathrm{GeV} \\
m_{H_3} &=& 542.946 \,\mathrm{GeV} & \; 
m_{H^\pm} &=& 223.758 \,\mathrm{GeV} \;.
\end{array}
\eeq
For the critical temperature $T_c$, the VEV $v_c$ at $T_c$ and $\xi_c$
we find for this parameter point
\beq
T_c &= 138.913\,\mathrm{GeV} \, , \; v_c = 139.274 \,\mathrm{GeV} \, ,
\; \xi_c = 1.0026 \;.
\eeq
The individual doublet VEVs $\bar{\omega}_1$ and $\bar{\omega}_2$ and the CP- and
charge-breaking VEVs $\bar{\omega}_{\text{CP}}$ and
$\bar{\omega}_{\text{CB}}$ at $T_c$ are  
\beq
\begin{array}{lcllcl}
\bar{\omega}_1(T_c) &=& 16.9487\,\mathrm{GeV} & \; 
\bar{\omega}_2(T_c) &=& 135.556\,\mathrm{GeV} \\
\bar{\omega}_{\text{CP}}(T_c) &=& 27.1021 \,\mathrm{GeV} & \; 
\bar{\omega}_{\text{CB}}(T_c) &=& 0 \,\mathrm{GeV} \;.
\end{array}
\eeq
We observe in Fig.~\ref{fig:PlotGen} (a) the jump for the symmetric
phase to a non-zero VEV with $v_c = 139.274$~GeV at $T_c=138.913$~GeV
corresponding to a strong first order EWPT with $\xi_c$ just above 1,
$\xi_c=1.0026$. For the chosen parameter point with $\tan\beta \approx
7$, the non-zero doublet VEV $\bar{\omega}_2$ is much larger than 
$\bar{\omega}_1$, {\it
  cf}~Fig.~\ref{fig:PlotGen} (b). Their squared sum approaches $v(T=0)=
\sqrt{\bar{\omega}_1^2+ \bar{\omega}_2^2} = 246.22$~GeV at zero
temperature. As can be inferred from
Fig.~\ref{fig:PlotGen} (c) and (d), at $T_c$, a
CP-violating phase $\bar{\omega}_3 \ne 0$ is generated
spontaneously at the EWPT. The non-physical charge-breaking VEV
$\bar{\omega}_{\text{CB}}$ on 
the other hand remains zero throughout the whole scanned temperature
interval, as it should,  {\it cf}~Fig.~\ref{fig:PlotGen} (d). 

\section{How to add a New Model \label{sec:newmodels}} 
In this section we describe how a new model can be added to the
program. To illustrate this, we have generated the template class 
{\tt ClassTemplate.cpp}, located in the directory {\tt
  BSMPT/src/models/}, in which the functions (according to the given comments)
have to be edited. The functions to be modified are
\begin{itemize}
\item[] {\tt Class\_Template}
\item[] {\tt ReadAndSet}
\item[] {\tt addLegendCT}
\item[] {\tt addLegendTemp}
\item[] {\tt addLegendTripleCouplings}
\item[] {\tt addLegendVEV}
\item[] {\tt set\_gen}
\item[] {\tt set\_CT\_Pot\_Par}
\item[] {\tt write}
\item[] {\tt TripleHiggsCouplings}
\item[] {\tt calc\_CT}
\item[] {\tt MinimizeOrderVEV}
\item[] {\tt SetCurvatureArrays}
\item[] {\tt CalculateDebyeSimplified}
\item[] {\tt VTreeSimplified}
\item[] {\tt VCounterSimplified}
\item[] {\tt Debugging}\footnote{In fact,
  the function {\tt Debugging} is not used by any of the programs and
  is provided only for the user to perform some checks.}
\end{itemize}
Furthermore, the constant of the new
model with which it is selected by the program (through {\it
  Model}) has to be defined in the file {\tt
  IncludeAllModels.h}. After doing so, in the file \\ {\tt
  IncludeAllModels.cpp} the corresponding entry in the
function {\tt Fchoose} has to be added, and the file
needs to be extended to include the new model. 
Additionally, in {\tt ClassTemplate.h} the parameters of the model
have to be declared. All these files are also located in {\tt BSMPT/src/models/}. 

\subsection{Example}
As example we take a model with one scalar particle $\phi$ which
develops a VEV $v$, couples to one fermion $t$ with the Yukawa
coupling $y_t$, and to one gauge boson $A$ with the gauge coupling
$g$. The relevant pieces of the Lagrangian are given by  ($\Phi = \phi
+ v$)
\begin{align}
-\mathcal{L}_S &= \frac{m^2}{2} \left(\phi+v\right)^2 +
                 \frac{\lambda}{4!}
                 \left(\phi+v\right)^4 \label{eq:templatetree} \\ 
-\mathcal{L}_F &= y_t t_L t_R \left(\phi + v\right) \\
\mathcal{L}_G &= g^2A^2\left(\phi+v\right)^2 \,.
\end{align} 
We therefore have $i,j,k,l = 1 , I,J = 1,2, a,b = 1$ for the tensors
defined in Eqs.~(\ref{Eq:LS_Classical}), (\ref{Eq:LF_Classical}) and
(\ref{Eq:LG_Classical}). Here $I,J=1,2$ corresponds to $t_L$ and
$t_R$, the left- and right-handed projections of the fermion $t$. 
The tensors are given by 
\begin{align}
L^i &= \left.\partial_{v} \left(-\mathcal{L}_S \right)\right|_{\phi=0,v=0} = 0 \\
L^{ij} &= \left.\partial_{v}^2 \left(-\mathcal{L}_S \right)\right|_{\phi=0,v=0} = m^2 \\
L^{ijk} &=  \left.\partial_{v}^3 \left(-\mathcal{L}_S \right)\right|_{\phi=0,v=0} = 0 \\
L^{ijkl} &=  \left.\partial_{v}^4 \left(-\mathcal{L}_S
           \right)\right|_{\phi=0,v=0} = \lambda \\ 
Y^{IJk} &= \begin{cases} 0 & I = J \; (I,J=t_L,t_R)\\
y_t & I \ne J \; (I,J=t_L,t_R) \end{cases} \\
G^{abij} &= \partial_A^2 \partial_v^2 \left( \mathcal{L}_G \right) =
           4g^2 \;.
\end{align}
The counterterm potential, given by Eq.~(\ref{Eq:CTPot}), reads
\begin{align}
V^{\text{CT}} &= \frac{\delta m^2}{2} \left(\phi+v\right)^2 + \frac{\delta
          \lambda}{4!} \left(\phi+v\right)^4 + \delta T \left(\phi+
          v\right) \;. \label{eq:templatecounterpot}
\end{align}
Application of Eqs.~(\ref{Eq:Renorm1}) and (\ref{Eq:Renorm2}) yields
\begin{align}
\delta T + v \delta m^2  + \frac{1}{6} v^3 \delta \lambda
 &= -\left.\partial_{\phi} V^{CW}\right|_{\phi=0} \\ 
\delta m^2 + \frac{v^2}{2}\delta \lambda &= -\left. \partial^2_{\phi}
                                           V^{CW}\right|_{\phi=0} \;.
\end{align}
The system of equations is overconstrained. Choosing
\beq
\delta T = t \;, \quad \mbox{with} \quad t \in \mathbb{R} \;,
\label{eq:cttad}
\eeq
we get
\beq
\delta \lambda &=& \frac{3t}{v^3} + \frac{3}{v^3} \left(\left.\partial_{\phi}
  V^{\CW}\right|_{\phi=0}\right) - \frac{3}{v^2} \left(\left. \partial^2_{\phi}
V^{\CW}\right|_{\phi=0} \right) \label{eq:ctlambda} \\
\delta m^2 &=& -\frac{3}{2v} \left( \left.\partial_{\phi}
  V^{\CW}\right|_{\phi=0}\right) + \frac{1}{2} \left(\left. \partial^2_{\phi}
V^{\CW}\right|_{\phi=0}\right) - \frac{3t}{2v} \;. \label{eq:ctms}
\eeq
To implement this model, several files need to be changed, as
described in the following. 

\subsubsection{IncludeAllModels.h, IncludeAllModels.cpp,
  ClassTemplate.h}
In {\tt IncludeAllModels.h} the constant with which the program 
selects the new model has to be set. Please make sure that the new
model number is not used already by an implemented model. The program
would then not know which model to select. Choosing for the template
model {\it e.g.}~5, this results in adding the line \\[0.1cm] 
\begin{kasten*}{}
\vspace*{-0.2cm}
const int C\_ModelTemplate=5;
\end{kasten*} \\
This model selection then has to be entered in {\tt
  IncludeAllModels.cpp} by adding to the function {\tt Fchoose} the
line 
\\[0.1cm]
\begin{kasten*}{}
\vspace*{-0.2cm}
else if(choice == C\_ModelTemplate) \\
	\{ \\
		return std::unique\_ptr$<$Class\_Potential\_Origin$>$ \{ new Class\_Template \}; \\
	\} \\
\end{kasten*} \\
In {\tt IncludeAllModels.cpp} the new model is included by adding the
line \\[0.1cm]
\begin{kasten*}{}
\vspace*{-0.2cm}
\#include ``ClassTemplate.h''
\end{kasten*} \\
In {\tt ClassTemplate.h} the variables for the potential
and for the remaining Higgs coupling parameters as well as for the
counterterm constants have to 
be added, \\[0.1cm]
\begin{kasten*}{}
double ms, lambda, dms, dlambda, dT, yt, g;
\end{kasten*} \\
Here 'ms' denotes the mass parameter squared, $m^2$, and 'dms',
'dlambda' are the counterterms $\delta m^2$, $\delta \lambda$. 

\subsubsection{ClassTemplate.cpp}
We will not describe here in detail every function in {\tt Class\_Template.cpp}	
that can be modified as the functions are commented in the code. Instead, we
briefly describe here the most essential parts. \s

\paragraph*{Class\_Template()}
The numbers of Higgs particles, potential parameters,
counterterms and VEVs have to be specified in the constructor 
{\tt Class\_Template}() and the variable 'Model' has to be set to
the selected model. In our simple example, this is \\[0.1cm]
\begin{kasten*}{}
	\vspace*{-0.6cm}
\begin{align}
\text{Model} &= \text{C\_ModelTemplate}; \\
\text{NNeutralHiggs} &= 1; \\
\text{NChargedHiggs} &= 0; \\
\text{nPar} &= 2; \\
\text{nParCT} &= 3; \\
\text{nVEV} &= 1; \\
\text{NHiggs} &= \text{NNeutralHiggs} + \text{NChargedHiggs};
\end{align}	 
	\vspace*{-0.6cm}
\end{kasten*} \\
When you implement a new model that is not called {\tt Template} but
{\it e.g.}~{\tt NewModel}, please make sure to replace in the corresponding {\tt .h} and
{\tt .cpp} files the name {\tt Class\_Template} by the name of the
newly implemented class. This means that {\tt
  Class\_Template} has to be replaced by {\tt Class\_NewModel} wherever it appears. 

\paragraph*{ReadAndSet(const std::string\& linestr, std::vector$<$double$>$\& par)}
In this function the input parameters of the model are read into the
vector 
'par'. Each line in the input file corresponds to a new parameter
point. The line to be read in is given by the string 
'linestr'. Via 'std::stringstream' the parameters of each line are
read into double variables. In our template model the input file would
contain the parameters 'ms' and 'lambda' so that in the
program it would look like this: \\[0.1cm]
\begin{kasten*}{}
	\vspace*{-0.6cm}
	\begin{align*}
		&\text{std::stringstream ss(linestr);} \\
		&\text{double tmp;} \\
		&\text{double lms,llambda;}\\
		&\text{for(int k=1;k}<=\text{2;k++)} \\
		&\text{\{}\\
	      &\hspace*{2em}\text{ss}>>\text{tmp;}\\
	      &\hspace*{2em}\text{if(k==1) lms = tmp; }\\
	      &\hspace*{2em}\text{else if(k==2) llambda = tmp;}\\
		&\text{\}}\\
		&\text{par[0] = lms;}\\
		&\text{par[1] = llambda;}
	\end{align*}
	\vspace*{-0.6cm}
\end{kasten*}

\paragraph*{set\_gen(const std::vector$<$double$>$\& par)}
Here, the potential parameters are set from the
vector 'par' read in with the function 
{\tt ReadAndSet(std::string linestr, double* par)}, as well as the
coupling parameters. In our sample model, the gauge coupling $g$ is
given by the SM gauge coupling, and the Yukawa coupling $y_t$  is given in
terms of the SM VEV and the top quark mass. The SM gauge coupling, the
SM VEV and the top quark mass are defined in {\tt \$BSMPT/src/models/SMparam.h}.
With the parameters 'ms' and 'lambda' this would then look like:
\\[0.1cm]
\begin{kasten*}{}
	\vspace*{-0.6cm}
	\begin{align*}
		\text{ms} &= \text{p}[0]; \\
        \text{lambda} &= \text{p}[1];\\
         \text{g} &= \text{C\_g}; \\
         \text{yt} &= \text{std::sqrt(2)/C\_vev0 * C\_MassTop};
	\end{align*}
	\vspace*{-0.6cm}
\end{kasten*} \\
More complicated Higgs sectors require
additional parameters. Furthermore, you can set here the potential
parameters that are not read in from the input parameters but are
calculated through the tree-level minimum conditions, like $m_{11}^2,
m_{22}^2$ and $\mbox{Im}(m_{12}^2)$ in the C2HDM {\it e.g.} 
This function is also used to define the vectors {\tt vevTree}
and {\tt vevTreeMin}. The former vector refers to the complete field
configuration appearing in the effective potential. The size of the
vector is hence given by $n_{\text{Higgs}}$ ({\it
  cf.}~Sec.~\ref{sec:notation}). For the (C)2HDM {\it 
  e.g.}, we would have $n_{\text{Higgs}}=8$ corresponding to the eight
real fields $\phi_i$ in 
Eq.~(\ref{eq:vec2hdm}) (Eq.~(\ref{eq:vecc2hdm})). The vector {\tt
  vevTreeMin} corresponds to the VEVs at $T=0$. Its size is given by
the field configurations 
that develop a VEV, {\it i.e.}~$n_v$ ({\it
  cf.}~Sec.~\ref{sec:counterpot}). This would be $n_v =4$ 
in the (C)2HDM, corresponding to the four VEVs $v_1,v_2,v_{\text{CP}}$
and $v_{\text{CB}}$. In our
simple template model $n_{\text{Higgs}}$ and $n_v$ coincide resulting
in two vectors {\tt vevTree} and {\tt vevTreeMin }of dimension 1
each. The value of {\tt vevTreeMin} is given by the SM VEV 'C\_vev0' that is
hard-coded in the program. In our sample model it would look like this: \\[0.1cm] 
\begin{kasten*}{}
\vspace*{-0.6cm}
\begin{align*}
&\text{vevTreeMin.resize(nVEV)} \,;\\
&\text{vevTreeMin}[0]= \text{C\_vev0} \,;\\
&\text{vevTree.resize(NHiggs)}\,;\\
&\text{MinimizeOrderVEV(vevTreeMin,vevTree)}\,;
\end{align*}
\vspace*{-0.6cm}
\end{kasten*}

Additionally, the $\overline{\mbox{MS}}$ renormalization scale can be
changed here through the command \\[0.1cm]
\begin{kasten*}{}
\vspace*{-0.6cm}
\begin{align*}
	\text{scale} &= \text{mu};
\end{align*}
\vspace*{-0.6cm}
\end{kasten*}\\
Here 'mu' is the chosen value in GeV for the renormalization
scale. The default value is 'mu = C\_vev0', {\it i.e.}~the EW VEV. 

\paragraph*{MinimizeOrderVEV(const std::vector$<$double$>$\&
  vevminimizer, \\
		std::vector$<$double$>$\& vevFunction)}
Whenever we deal with the Higgs potential in the calculation, the
dimension of the vector describing the fields is $n_{\text{Higgs}}$. Not
all of these fields develop VEVs, however, so that the vector used in
the minimizer only has dimension $n_v$. 
The function {\tt MinimizeOrderVEV} is used to convert the resulting
vector from the minimizer to the vector with the $n_{\text{Higgs}}$
entries. In order to do so the field(s) that develop(s) VEV(s) have to 
be selected. In the template model we have only one field and it develops a
VEV so that we simply have to set \\[0.1cm]
\begin{kasten*}{}
\vspace*{-0.6cm}
\begin{align*}
\text{VevOrder[0]} &= 0\,;
\end{align*}
\vspace*{-0.6cm}
\end{kasten*} \\
In a more complex model with {\it e.g.}~two fields where only one of
them develops a VEV, one would have to set '$\text{VeVOrder}[0]=0$' if 
the field developing the VEV is in the first entry of the vector
describing the fields, and '$\text{VeVOrder}[0]=1$' if it is the field
in the second entry.  

\paragraph*{SetCurvatureArrays()}
The tensors of the Lagrangian of the new model have to be implemented
in the function {\tt SetCurvatureArrays()}. The
notation is \\[0.1cm]
\begin{kasten*}{}
\vspace*{-0.6cm}
\begin{align*}
\text{Curvature\_Higgs\_L1}[i] &= L^i \\
\text{Curvature\_Higgs\_L2}[i][j] &= L^{ij} \\
\text{Curvature\_Higgs\_L3}[i][j][k] &= L^{ijk} \\
\text{Curvature\_Higgs\_L4}[i][j][k][l] &= L^{ijkl} \\
\text{Curvature\_Gauge\_G2H2}[a][b][i][j] &= G^{abij} \\
\text{Curvature\_Quark\_F2H1}[I][J][k] &= Y^{IJk}\,.
\end{align*}
\vspace*{-0.6cm}
\end{kasten*} \\
Technically, one could use 'Curvature\_Quark\_F2H1' to store all quarks
and leptons there, but as they do not mix the program provides besides
'Curvature\_Quark\_F2H1' where $I,J$ run over all quarks, also the 
structure 'Curvature\_Lepton\_F2H1[I][J][k]' where $I,J$ run over all
leptons. For our example this would look like \\[0.1cm]
\begin{kasten*}{}
\vspace*{-0.6cm}
\begin{align*}
\text{Curvature\_Higgs\_L1}[0] &= 0 ;\\
\text{Curvature\_Higgs\_L2}[0][0] &= \text{ms} ;\\
\text{Curvature\_Higgs\_L3}[0][0][0] &= 0 ;\\
\text{Curvature\_Higgs\_L4}[0][0][0][0] &= \text{lambda} ;\\
\text{Curvature\_Gauge\_G2H2}[0][0][0][0] &= \text{4*std::pow(g,2)} ;\\
\text{Curvature\_Quark\_F2H1}[0][0][0] &= 0 ;\\
\text{Curvature\_Quark\_F2H1}[1][0][0] &= \text{yt} ;\\
\text{Curvature\_Quark\_F2H1}[0][1][0] &= \text{yt} ;\\
\text{Curvature\_Quark\_F2H1}[1][1][0] &= 0 ;
\end{align*}
\vspace*{-0.6cm}
\end{kasten*}{}

\paragraph*{set\_CT\_Pot\_Par(const std::vector$<$double$>$\& par)}
For the use of the counterterms, the corresponding vectors for the
counterterm potential have to be set. They are named 
  'Curvature\_Higgs\_CT\_L1', 'Curvature\_Higgs\_CT\_L2',
  'Curvature\_Higgs\_CT\_L3' and \\ 'Curvature\_Higgs\_CT\_L4' and
  defined analogously to 'Curvature\_Higgs\_L1' to \\ 'Curvature\_Higgs\_L4'.

\paragraph*{calc\_CT( std::vector$<$double$>$\& par)}
The counterterms are computed numerically in the function  {\tt
  calc\_CT( std::vector$<$double$>$\& par)}. To do so, the user has to implement the
formulae for the counterterms that were derived beforehand
analytically in terms of the derivatives of the Coleman-Weinberg
potential, {\it cf.}~Eqs.~(\ref{eq:cttad}), (\ref{eq:ctlambda}) and
(\ref{eq:ctms}) for our template 
model. The derivatives of $V^{\CW}$ are provided by the program
through the function calls {\tt WeinbergFirstDerivative} and {\tt
  WeinbergSecondDerivative}. In detail, to calculate the counterterms $\delta m^2$,
$\delta \lambda$ and $\delta T$ of the template model, the following
steps have to be performed: 
\begin{itemize}
\item To calculate the first and second derivative of the
  Coleman-Weinberg potential call \\[0.1cm]
\begin{kasten*}{}
\vspace*{-0.2cm}
    std::vector$\langle$double$\rangle$ WeinbergNabla,WeinbergHesse; \\
    WeinbergFirstDerivative(WeinbergNabla); \\
    WeinbergSecondDerivative(WeinbergHesse);
\end{kasten*} \\ 
and to save it in a vector and matrix class use \\[0.1cm]
\begin{kasten*}{}
\vspace*{-0.2cm}
    	VectorXd NablaWeinberg(NHiggs); \\
	MatrixXd HesseWeinberg(NHiggs,NHiggs); \\
	for(int i=0;i$<$NHiggs;i++) \\
	\{ \\
        \hspace*{2em} NablaWeinberg[i] = WeinbergNabla[i]; \\
        \hspace*{2em} for(int j=0;j$<$NHiggs;j++) \\
        \hspace*{2em}\{\\
        	\hspace*{4em} HesseWeinberg(i,j) = WeinbergHesse.at(j*NHiggs+i); \\
        \hspace*{2em} \} \\
       \}
\end{kasten*}
\item Implement the previously derived formulae for the
  counterterms. In our example, these are
  Eqs.~(\ref{eq:cttad}), (\ref{eq:ctlambda}) and (\ref{eq:ctms}),
  where we set $t =0$, \\[0.1cm]
\begin{kasten*}{}
\vspace*{-0.6cm}
\begin{align*}
    		\text{dT} =& 0 ; \\
      		\text{dlambda} =& \text{3.0/std::pow(C\_vev0,3) *
                                 \text{NablaWeinberg[0]}} \\ &- \text{
                          3.0/std::pow(C\_vev0,2) *
                                 \text{HesseWeinberg(0,0)}} ;\\
      		\text{dms} =& \text{-3.0/(2*C\_vev0) * \text{NablaWeinberg[0]} + 1.0/2.0
                      *\text{HesseWeinberg(0,0)}}  ;
\end{align*}
\vspace*{-0.6cm}
\end{kasten*} 

\item Insert the parameters in the vector 'par', \\[0.1cm]
\begin{kasten*}{}
\vspace*{-0.6cm}
\begin{align*}
      	\text{par}[0] &= \text{dT} ; \\
      	\text{par}[1] &= \text{dms} ; \\
      	\text{par}[2] &= \text{dlambda} ;      	
\end{align*}
\vspace*{-0.6cm}
\end{kasten*}
\item Finally call  \\[0.1cm] 
\begin{kasten*}{}
\vspace*{-0.6cm}
	\begin{align*}
		\text{set\_CT\_Pot\_Par(par);}
	\end{align*}
\vspace*{-0.6cm}
\end{kasten*} \\
so that everything is set correctly. 
\end{itemize}
Afterwards, the values for dT, dmS and dlambda are set from the vector
'par' by the function {\tt set\_CT\_Pot\_Par(const std::vector$<$double$>$\& par)}.
  
\paragraph*{TripleHiggsCouplings()}
This function provides the trilinear loop-corrected
Higgs self-coup\-lings as obtained from the effective potential. They
are calculated from the third derivative of the Higgs potential
with respect to the Higgs fields in the gauge basis and then rotated
to the mass basis. Since the Higgs fields are ordered by mass,
{\it i.e.}~we have ascending indices with ascending mass, and the mass
order can change with each parameter point, this implies that for each
parameter point the indices of the vector 
containing the trilinear Higgs coupling would refer to different Higgs
bosons. Therefore, it is necessary to order the Higgs bosons in the
mass basis irrespective of the mass order. This order is defined
through the vector {\tt HiggsOrder(NHiggs)}. \\[0.1cm]
\begin{kasten*}{}
\vspace*{-0.3cm}
	for(int i=0;i$<$NHiggs;i++) \\
	\{ \\
        \hspace*{2em} HiggsOrder[i]= value; \\
     \}
\end{kasten*}  \\
The number 'value' is defined by the user according
to the ordering that this desired in the mass basis. Thus {\tt
  HiggsOrder}$[0]=5$ {\it e.g.}~would assign the 6th lightest particle to the first
position. The particles can be selected through the mixing matrix elements.
  
\paragraph*{addLegendTripleCouplings()}
All the following functions {\tt addLegend...} extend the legends of
the output files by certain variables. The function {\tt
  addLegendTripleCouplings} extends the legend by the column names for
the trilinear Higgs couplings derived from the tree-level, the
  counterterm and the Coleman-Weinberg potential. In order to do so,
  the user first has to make sure to define the names of the Higgs
  particles of the model in the vector 'particles'. In our
    model we only have one Higgs 
  particle that we call $H$ and hence set 'particles[0]="H";'. 

\paragraph*{addLegendTemp()}
Here the column names for $T_c$, $v_c$ and the VEVs are added to the
legend. The order should be $T_c$,  $v_c$ and then the names of the individual
VEVs. These VEVs have to be added in the same order as given
in the function {\tt MinimizeOrderVEV}.

\paragraph*{addLegendVEV()}
This function adds the column names for the VEVs
that are given out. The order has to be the same as given in the
function {\tt MinimizeOrderVEV}.

\paragraph*{addLegendCT()}
In this function, the legend for the counterterms is added. The order of
the counterterms has to be same as the one set in the function {\tt
  set\_CT\_Pot\_Par(par)}. 
  
\paragraph*{VTreeSimplified, VCounterSimplified}
The functions \\ {\tt VTreeSimplified(const std::vector$<$double$>$\&
  v)} and \\
{\tt VCounterSimplified(const std::vector$<$double$>$\& v)} can be used to
explicitly implement the formulae for the tree-level and counterterm potential in
terms of the classical fields $\omega$, in our
example these are Eqs.~(\ref{eq:templatetree}) and 
Eq.~(\ref{eq:templatecounterpot}), respectively, with  
$\phi=0$ and $v \equalhat \omega$. Implementing these
  may improve the runtime of the programs. An example is given in the template class.
  
\paragraph*{CalculateDebyeSimplified(), CalculateDebyeGaugeSimplified()}
The functions \\ {\tt CalculateDebyeSimplified()} and {\tt
  CalculateDebyeGaugeSimplified()} can be used to implement
explicit formulae for the daisy corrections to the masses of the scalars,
{\it cf.}~Eq.~(\ref{eq:pis}), and gauge bosons,
Eq.~(\ref{eq:pigauge}), respectively. This is done by setting the vectors
'DebyeHiggs' and 'DebyeGauge' and finishing the function with a return true statement. 

\paragraph*{write()}
The function {\tt write()} can be used to give a terminal output of
the potential parameters. For our example this would be  \\[0.1cm]
\begin{kasten*}{}
	\begin{align*}
	\text{std::cout} &<< \text{"The parameters are : " $<<$ std::endl;}\\
	\text{std::cout} &<< \text{"lambda = " $<<$ lambda $<<$ std::endl}\\
	&<< \text{"\textbackslash tm\^{}2 = " $<<$ ms $<<$ std::endl;}\\
	\text{std::cout} &<< \text{"The counterterm parameters are : " $<<$ std::endl;}\\
	\text{std::cout} & << \text{ "dT = "$<<$ dT $<<$ std::endl}\\
				&<< \text{"dlambda = " $<<$ dlambda $<<$ std::endl}\\
			&<< \text{"dm\^{}2 = "$<<$ dms $<<$ std::endl;}\\
	\text{std::cout} &<< \text{"The scale is given by mu = " $<<$ scale $<<$ " GeV " $<<$ std::endl;}
	\end{align*}
\end{kasten*} 

\section{Summary \label{sec:concl}}
We have presented the {\tt C++} package {\tt BSMPT} for the
investigation of electroweak baryogenesis in extended Higgs sectors
beyond the SM. The package calculates the loop-corrected effective
potential at finite temperature including daisy resummations of the
bosonic masses. It can be used for the computation of the VEV as a
function of the temperature and in particular for the determination of
$\xi_c = v_c/T_c$ which is related to the strength of the phase
transition. Furthermore, the loop-corrected trilinear Higgs self-couplings
are given out, allowing to investigate the interplay between
successful baryogenesis and the required size on the Higgs
self-interactions. The chosen 'on-shell' renormalization scheme
enables efficient scans in the parameter scans of the models and
allows for the analysis of the connection between collider
phenomenology and successful baryogenesis, so that a link between
collider phenomenology and cosmology can be made. The already 
implemented models are the CP-conserving and CP-violating 2HDMs and
the N2HDM. The program
structure supports the implementation of new models, and we have
illustrated with the help of a toy model how this can be done. With
our new tool at hand, it is easy to further investigate the possibility
of baryogenesis in new physics models, the possible spontaneous
generation of CP-violating phases and make further links between
collider observables and phenomena like {\it e.g.}~gravitational waves. The
program is constantly updated to include new phenomenologically
interesting models. We are grateful for suggestions. 

\section*{Acknowledgements}
The authors thank Jonas M\"uller, Jonas Wittbrodt and Alexander
Wlotzka for many useful discussions and assistance during the
debugging process. They furthermore thank Jonas Wittbrodt for the careful reading of
the manuscript. PB acknowledges financial support by the “Karlsruhe
School of Elementary Particle and Astroparticle Physics: Science and
Technology (KSETA)”. 

\vspace*{1cm}

\end{document}